\documentclass{aa}

\usepackage{graphicx}
\usepackage{txfonts}
\usepackage{hyperref}
\usepackage{subfig}

\graphicspath{{figures/}}

\hypersetup{
  colorlinks=true,
  linkcolor=blue,
  urlcolor=blue,
  citecolor=blue
}

\newcommand{\toprule}{\hline\hline}
\newcommand{\midrule}{\hline}
\newcommand{\bottomrule}{\hline}
\newcommand{\planck}{{\sl Planck}\xspace}
\newcommand{\nv}{\hat{\bf n}}

\defcitealias{Alonso:2020}{A21}
\defcitealias{Hale:2023}{H23}

\newcommand{\sigmafid}{{\sigma_8 = 0.75^{+0.05}_{-0.04}}}
\newcommand{\sigmatest}{{\sigma_8 = 0.82^{+0.08}_{-0.07}}}

\newcommand{\biasd}{{b_{g,D} = 1.41 \pm {0.06}}}
\newcommand{\biasdtest}{{b_{g,D} = 1.40 \pm 0.07}}
\newcommand{\biasval}{{b(z=0.82) = 2.34 \pm {0.10}}}

\newcommand{\zfid}{{0.05 \pm 0.01}}
\newcommand{\rfid}{{0.20 \pm 0.03}}
\newcommand{\afid}{{4.9 \pm 0.1}}

\newcommand{\ztest}{{0.04 \pm 0.01}}
\newcommand{\rtest}{{0.17 \pm 0.03}}
\newcommand{\atest}{{5.0 \pm 0.1}}

\newcommand{\sigmaCgglow}{{17.9}}
\newcommand{\sigmaCgghigh}{{34.6}}
\newcommand{\sigmaCgklow}{{23.1}}
\newcommand{\sigmaCgkhigh}{{26.6}}

\begin{document} 

\title{Cosmology from LOFAR Two-metre Sky Survey Data Release 2: Cross-correlation with the cosmic microwave background}

\titlerunning{Cosmology from LOFAR LoTSS DR2: Cross-correlation with the CMB}
\authorrunning{Nakoneczny et al.}

\author{
    S.J.~Nakoneczny\inst{\ref{caltech},\ref{ncbj}} \and
    D.~Alonso\inst{\ref{oxford}} \and
    M.~Bilicki\inst{\ref{cft}} \and
    D.J.~Schwarz\inst{\ref{bielefeld}} \and
    C.L.~Hale\inst{\ref{edinburgh}} \and
    A.~Pollo\inst{\ref{ncbj},\ref{uj}} \and
    C.~Heneka\inst{\ref{heidelberg}} \and
    P.~Tiwari\inst{\ref{GTIIT}} \and
    J.~Zheng\inst{\ref{bielefeld}} \and
    M.~Brüggen\inst{\ref{hamburg}} \and    
    M.J.~Jarvis\inst{\ref{oxford},\ref{westerncape}} \and
    T.W.~Shimwell\inst{\ref{astron},\ref{leiden}}
}

\institute{
    Division of Physics, Mathematics and Astronomy, California Institute of Technology, 1200 E California Blvd, Pasadena, CA 91125
    \label{caltech}
    \and
    Department of Astrophysics, National Centre for Nuclear Research, Pasteura 7, 02-093 Warsaw, Poland
    \label{ncbj}
    \and
    Sub-department of Astrophysics, University of Oxford, Denys Wilkinson Building, Keble Road, Oxford, OX1 2DL, UK
    \label{oxford}
    \and
    Center for Theoretical Physics, Polish Academy of Sciences, al. Lotnik\'{o}w 32/46, 02-668 Warsaw, Poland \label{cft} \and
    Fakult\"at f\"ur Physik, Universit\"at Bielefeld, Postfach 100131, 33501 Bielefeld, Germany
    \label{bielefeld}
    \and
    Institute for Astronomy, School of Physics and Astronomy, University of Edinburgh, Royal Observatory Edinburgh, Blackford Hill, Edinburgh, EH9 3HJ, UK \label{edinburgh}
    \and
    Astronomical Observatory of the Jagiellonian University, 31-007 Cracow, Poland
    \label{uj}
    \and
    Institute for Theoretical Physics, Heidelberg University, Philosophenweg 16, 69120 Heidelberg, Germany
    \label{heidelberg}
    \and 
    Department of Physics, Guangdong Technion - Israel Institute of Technology, Shantou, Guangdong 515063, P.R. China
    \label{GTIIT}
    \and 
    Hamburg Observatory, University of Hamburg, Gojenbersgweg 112, 21029 Hamburg, Germany
    \label{hamburg}
    \and 
    Department of Physics, University of the Western Cape, Bellville 7535, South Africa
    \label{westerncape}
    \and 
    ASTRON, Netherlands Institute for Radio Astronomy, Oude Hoogeveensedijk 4, 7991 PD, Dwingeloo, The Netherlands
    \label{astron}
    \and
    Leiden Observatory, Leiden University, PO Box 9513, NL-2300 RA Leiden, The Netherlands
    \label{leiden}
}

\authorrunning{S.J.~Nakoneczny et al.}

\offprints{S.J.~Nakoneczny, \email{\url{nakonecz@caltech.edu}}.}

\abstract
{}
{We combined the LOw-Frequency ARray (LOFAR) Two-metre Sky Survey (LoTSS) second data release (DR2) catalogue with gravitational lensing maps from the cosmic microwave background (CMB) to place constraints on the bias evolution of LoTSS-detected radio galaxies, and on the amplitude of matter perturbations.}
{We constructed a flux-limited catalogue from  LoTSS DR2, and analysed its harmonic-space cross-correlation with CMB lensing maps from \planck, $C_\ell^{g\kappa}$, as well as its auto-correlation, $C_\ell^{gg}$. We explored the models describing the redshift evolution of the large-scale radio galaxy bias, discriminating between them through the combination of both $C_\ell^{g\kappa}$ and $C_\ell^{gg}$. Fixing the bias evolution, we then used these data to place constraints on the amplitude of large-scale density fluctuations, parametrised by $\sigma_8$.}
{We report the significance of the $C_\ell^{g\kappa}$ signal at a level of \sigmaCgkhigh$\sigma$. We determined that a linear bias evolution of the form $b_g(z) = b_{g,D} / D(z)$, where $D(z)$ is the growth rate, is able to provide a good description of the data, and we measured $\biasd$ for a sample that is flux limited at $1.5\,{\rm mJy}$, for scales $\ell < 250$ for $C_\ell^{gg}$, and $\ell < 500$ for $C_\ell^{g\kappa}$. At the sample's median redshift, we obtained $\biasval$. Using $\sigma_8$ as a free parameter, while keeping other cosmological parameters fixed to the \planck values, we found fluctuations of $\sigmafid$. The result is in agreement with weak lensing surveys, and at $1\sigma$ difference with \planck CMB constraints. We also attempted to detect the late-time-integrated Sachs-Wolfe effect with LOFAR data; however, with the current sky coverage, the cross-correlation with CMB temperature maps is consistent with zero. Our results are an important step towards constraining cosmology with radio continuum surveys from LOFAR and other future large radio surveys.}
{}


\maketitle

\section{Introduction}\label{sec:intro}

One of the main current goals of observational cosmology is constraining the history of structure growth as a way to pin down the different components that dominate the background expansion of the Universe at late times \citep{Huterer:2023}. To do so, one must investigate probes of the structure that, ideally, satisfy a number of criteria. They should cover a large enough patch of the Universe, accessing the cosmological scales, connect with fundamental quantities, such as the matter overdensity, and contain redshift information, allowing for an accurate reconstruction of the structure growth history. Unfortunately, virtually no cosmological probe, taken alone, is able to fulfill these requirements. Although weak gravitational lensing, measured from its effect on the shapes of background galaxies or on the cosmic microwave background (CMB) fluctuations \citep{2001PhR...340..291B}, is an unbiased tracer of the total matter fluctuations, it has a significantly lower raw statistical power and poorer ability to trace redshift evolution than measurements of galaxy clustering. The latter, in turn, is cursed by the problem of galaxy bias: the complicated relation between galaxy and matter overdensities. However, if reasonably accurate redshifts are available (be them spectroscopic or photometric), the growth of a structure can be reconstructed through redshift-space distortions \citep{Guzzo:2008, 2011MNRAS.415.2876B, 
  delaTorre:2013,
  2013MNRAS.436.3089B, 2015MNRAS.449..848H, 2016PASJ...68...38O, 2017A&A...604A..33P, 2021PhRvD.103h3533A}, or by combining galaxy clustering and weak lensing \citep{2002PhRvD..66h3515H, 
  delaTorre:2017,
  2018MNRAS.481.1133P, 2019JCAP...10..015W, 2020JCAP...05..047K, Heymans:2021, 2022JCAP...02..007W, 2021JCAP...10..030G, Alonso:2023}. The combination of different tracers of the large-scale structure is thus able to overcome their individual shortcomings and fulfill the requirements listed above. It is for this reason that multi-tracer large-scale structure analyses have now become one of the staples of late-Universe cosmology.

In this context, radio continuum surveys are an interesting and promising probe. Due to the large instantaneous field of view of modern low frequency radio interferometers, such surveys cover wide areas of the sky. Unencumbered by dust extinction, they are able to cover large swathes of the Universe. Their ability to recover clustering information on gigaparsec scales thus makes them potentially valuable for specific cosmological science cases, such as the search for primordial non-Gaussianity \citep{Ferramacho:2014, Alonso:2015, Gomes:2020}. Furthermore, radio continuum samples are dominated by active galactic nuclei (AGNs) and star-forming galaxies (SFGs), and thus their study can shed light on key processes in the formation and evolution of galaxies. In this sense, the clustering of radio sources has been used to place constraints on the properties of the different radio populations, making use of a variety of datasets, such as Sydney University Molonglo Sky Survey \citep[SUMSS, ][]{Blake:2004}, NRAO VLA Sky Survey \citep[NVSS, ][]{Blake:2002, Overzier:2003, Negrello:2006, Nusser:2015, Chen:2016}, Faint Images of the Radio Sky at Twenty-cm \citep[FIRST, ][]{Lindsay:2014}, Cosmic Evolution Survey at 3GHz \citep[COSMOS 3GHz, ][]{Hale:2018}, and TIFR GMRT Sky Survey \citep[TGSS, ][]{Dolfi:2019, Rana:2019}. In this work, we study the clustering of radio galaxies in the second data release of the LOw-Frequency ARray Two-metre Sky Survey \citep[LoTSS DR2][]{Shimwell:2022}, extending the analyses carried out making use of the first data release \citep[DR1,][]{Siewert:2020, Alonso:2020, Tiwari:2022}.

The dominant mechanism for radio emission is synchrotron radiation, which is characterised by a featureless, almost power-law spectrum \citep{1992ARA&A..30..575C}. The absence of bright emission lines or other sharp features in the radio spectrum, therefore, precludes redshift measurement. This gives rise to two key sources of uncertainty in the cosmological analysis of radio continuum surveys: the evolution of the main properties of the sample (e.g. galaxy bias and relative fractions of different source types) over the large range of redshifts covered by the sample, and the detailed description of the sample's redshift distribution.
  
As in the case of optical surveys, some of these uncertainties can be overcome or mitigated through the use of cross-correlations. For example, their cross-correlation with optical catalogues can be used to infer the clustering properties of the radio sample, and to constrain its redshift distribution via tomography \citep[e.g.][]{Menard:2013}. Our focus in this work is on the cross-correlation with maps of the CMB lensing convergence \citep{Planck:2020:lensing}. CMB lensing, sourced at $z\sim1100$, receives contributions from density inhomogeneities covering a wide range of redshifts, peaking at $z\sim2$. As such, it is an interesting tracer to cross-correlate with radio data, one of the few probes able to cover comparable volumes. In fact, the cross-correlation with radio data from the NVSS \citep{NVSS} was used to carry out the first detection of the CMB lensing signal \citep{Smith:2007}. Since then, this cross-correlation has been used for the benefit of both probes, for example as a way to measure the radio galaxy bias \citep{Allison:2015,Piccirilli:2023}, and in the context of delensing \citep{Namikawa:2015}. As the sensitivity of radio and CMB experiments increases, and as the statistical uncertainties of this cross-correlation decrease, the joint analysis of radio continuum and CMB lensing data becomes a more powerful tool for cosmological studies, which are able to not only constrain amplitude-like parameters, but also to discriminate between more nuanced details of the underlying astrophysical model. For example, the analysis of the first LoTSS data release in combination with CMB lensing data \citep[A21 hereafter]{Alonso:2020} showed that the galaxy auto-correlation and its cross-correlation with CMB lensing respond differently to changes in the galaxy bias and to the width of the redshift distribution, two effects that would otherwise be highly degenerate. The inclusion of CMB lensing data can therefore shed light on the main systematic uncertainties affecting continuum surveys as described above. In addition, it provides a way to measure the global amplitude of matter fluctuations.

In order to constrain the redshift distribution of the LoTSS DR2 radio sources, we made use of the cross-identifications of radio sources with multi-wavelength data in three LoTSS deep fields \citep{Tasse:2021, Sabater:2021,Kondapally:2021}. These allowed for the measurement of photometric redshifts for over 90 per cent of all deep field sources at flux densities above $1.5 \,{\rm mJy}$, of which about a quarter of them have a spectroscopic redshift as well \citep{Duncan:2021, Bhardwaj:2023}.

In this paper, we address the problems of galaxy bias within flux-limited radio samples and its redshift distribution, and the amplitude of density fluctuations as probed by radio data. To do so, we carried out a joint harmonic-space analysis of two radio samples extracted from the LoTSS DR2, defined by different flux and signal-to-noise cuts, together with maps of the CMB lensing convergence provided by the \planck collaboration. The analysis of the LoTSS sample closely follows the treatment described in the companion paper \citep[H23 hereafter]{Hale:2023}, focussed on constraints from the real-space galaxy auto-correlation. Taking advantage of the tools developed for that analysis, we also applied them to the cross-correlation between LoTSS and CMB primary anisotropies, in order to place constraints on the integrated Sachs-Wolfe (ISW) effect \citep[][]{Sachs:1967}.

This paper is structured as follows: Section \ref{sec:theory} lays out the theoretical background behind the auto- and cross-correlations we use here. Sections \ref{sec:data} and \ref{sec:method} present the datasets used in this work, as well as the methods used to analyse them. The measurements of the radio galaxy bias and the tentative constraints on $\sigma_8$ are presented in Section \ref{sec:results}. We discuss and summarise our results in Sections \ref{sec:discussion} and \ref{sec:conclusions}.

\section{Theory}\label{sec:theory}

Our main observable is a sky map of the galaxy surface density
\begin{equation}
    \delta_g(\nv) = \frac{N_g(\nv) - \bar{N_g}}{\bar{N}_g},
\end{equation}
where $\nv$ is a unit vector pointing along a line of sight, $N_g(\nv)$ is the number of galaxies along $\nv$ per unit solid angle, and $\bar{N}_g$ is the mean number of galaxies per unit solid angle. The projected overdensity is related to the three-dimensional galaxy overdensity $\Delta_g$ through \citep{1980lssu.book.....P}
\begin{equation}
    \delta_g(\nv) = \int_0^\infty dz\,p(z)\, \Delta_g(\chi(z)\nv, z),
\end{equation}
where $\chi(z)$ is the comoving radial distance, and $p(z)$ is the redshift distribution of the galaxy sample, normalised to 1 when integrated over $z$.

In addition to $\delta_g$, we will study maps of the CMB lensing convergence $\kappa(\nv)$, which quantifies the distortion in the trajectories of the CMB photons caused by the gravitational potential of the intervening matter structures \citep{Lewis:2006}, and is proportional to the divergence of the deflection in the photon arrival angle $\boldsymbol{\alpha}$: $\kappa\equiv -\nabla_{\nv} \cdot \boldsymbol{\alpha}/2$. As such, $\kappa$ is an unbiased tracer of the matter density fluctuations $\Delta_m(\boldsymbol{x}, z)$, and is related to them through:
\begin{equation}
    \kappa(\nv) = \int_0^{\chi_{LSS}} d\chi \frac{3H_0^2 \Omega_m}{2a(\chi)} \chi \frac{\chi_{LSS} - \chi}{\chi_{LSS}} \Delta_m(\chi \nv, z(\chi)),
\end{equation}
where $\Omega_m$ is the fractional matter density, $a = 1/(1+z)$ is the scale factor, $H_0$ is the Hubble constant today, and $\chi_{LSS}$ is the comoving distance to the surface of last scattering.

Finally, we will also consider the cross-correlation with CMB temperature anisotropies, which receives a contribution from the so-called Integrated Sachs-Wolfe effect \citep{Sachs:1967}. Caused by time-varying gravitational potentials at late times, the ISW leads to an additional temperature fluctuation of the form
\begin{equation}
 \left.\frac{\Delta T}{T}\right|_{\rm ISW}(\nv)=2\int_0^{\chi_{\rm LSS}}d\chi\,a\,\dot{\phi},
\end{equation}
where $\dot{\phi}$ is the derivative of the Newtonian potential with respect to cosmic time. In Fourier space, $\dot{\phi}$ can be related to the matter overdensity, assuming linear growth, via
\begin{equation}
  \dot{\phi}({\bf k},t)=-\frac{3H_0^2\Omega_m}{2a}\frac{H}{k^2}(f-1)\,\Delta_m({\bf k},t),
\end{equation}
where $f(a)\equiv d\log\Delta_m/d\log a$ is the `growth rate', which is scale-independent in the linear regime.

Consider a generic three dimensional field ($U$) projected onto a sphere with a kernel $W_u$
\begin{equation}
    u(\nv) = \int d \chi W_u(\chi) U(\chi \nv, z(\chi)).
\end{equation}
Any such projected quantity can be decomposed in terms of its spherical harmonic coefficients $u_{\ell m}$, the covariance of which with another field $V$ is the so-called angular power spectrum ($C_\ell^{uv}$). The angular power spectrum can be related to the power spectrum of the 3D fields $P_{UV}(k,z)$ through
\begin{equation}
\label{eq:c_ell}
    C_\ell^{uv} = \int \frac{d \chi}{\chi^2} W_u(\chi) W_v(\chi) P_{UV}\left(k_\ell(\chi), z(\chi)\right),
\end{equation}
where $P_{UV}(k, z)$ is the covariance of the Fourier coefficients of $U$ and $V$, and $k_\ell(\chi)\equiv(\ell+1/2)/\chi$. In this formalism, for the three fields under consideration (galaxy overdensity, CMB lensing convergence, and ISW), the radial kernels are given by
\begin{align}
  &W_g(\chi) = \frac{H(z)}{c} p(z), \label{eq:W_g}\\
  &W_\kappa(\chi) = f_\ell \frac{3H_0^2 \Omega_m}{2a} \chi \frac{\chi_{LSS} - \chi}{\chi_{LSS}} \Theta(\chi_{LSS} - \chi),\\
  &W_{\rm ISW}(\chi)=\frac{3\,H_0^2\Omega_m}{k_\ell^2}H(z)(1-f),
\end{align}
where $H(z)$ is the expansion rate, $\Theta(x)$ is the Heaviside function, $c$ is the speed of light, and $f_\ell$ is the scale-dependent prefactor given by
\begin{equation}
  f_\ell = \frac{\ell(\ell + 1)}{(\ell + 1/2)^2} = 1-\frac{1}{(2\ell+1)^2},
\end{equation}
which significantly differs from unity only for $\ell \lesssim 10$, and accounts for the fact that $\kappa$ is related to $\Delta_m$ through the angular Laplacian of the gravitational potential $\phi$. Eq.~\eqref{eq:c_ell} is only valid in the Limber approximation \citep{Limber:1953}, which holds when the extent of the radial kernels is much broader than the correlation scale of the matter fluctuations (which is the case for $W_g$, $W_\kappa$ and $W_{\rm ISW}$ in this work).

In order to use the galaxy distribution as a probe of structure, we need to define its relation to the matter overdensities. This implies developing models for the 3D power spectra $P_{gg}(k, z)$, and $P_{gm}(k, z)$, as well as the matter power spectrum $P_{mm}(k,z)$. In this work, we will assume a simple parametrisation, for which
\begin{equation}
  P_{gg}(k,z)=b_g^2(z) P_{mm}(k,z),\hspace{12pt} P_{gm}(k, z) = b_g(z)P_{mm}(k, z),
\end{equation}
where $b_g(z)$ is the linear bias function. The models used to describe the redshift distribution of our sample, and the redshift evolution of the bias are described in Section \ref{sec:method}.

To compute the linear matter power spectrum, we use the {\tt CAMB} Boltzmann solver \citep{Lewis:2000}, and estimate the non-linear power spectrum from it using {\tt HALOFIT} \citep{Smith:2003, Takahashi:2012}. All other theoretical calculations (e.g. Limber integrals) were carried out using the Core Cosmological Library \citep[{\tt CCL}, ][]{Chisari:2019}. Unless stated otherwise, we fix all cosmological parameters to the best-fit $\Lambda$CDM values found by 
\cite{Planck:2020:overview}: $\Omega_c = 0.26503$, $\Omega_b = 0.04939$, $h = 0.6732$, $\sigma_8 = 0.8111$, $n_s = 0.96605$.

\section{Data}\label{sec:data}

\subsection{LoTSS DR2}\label{ssec:data.lotss}
\begin{figure*}
    \centering
    \includegraphics[width=0.48\textwidth]{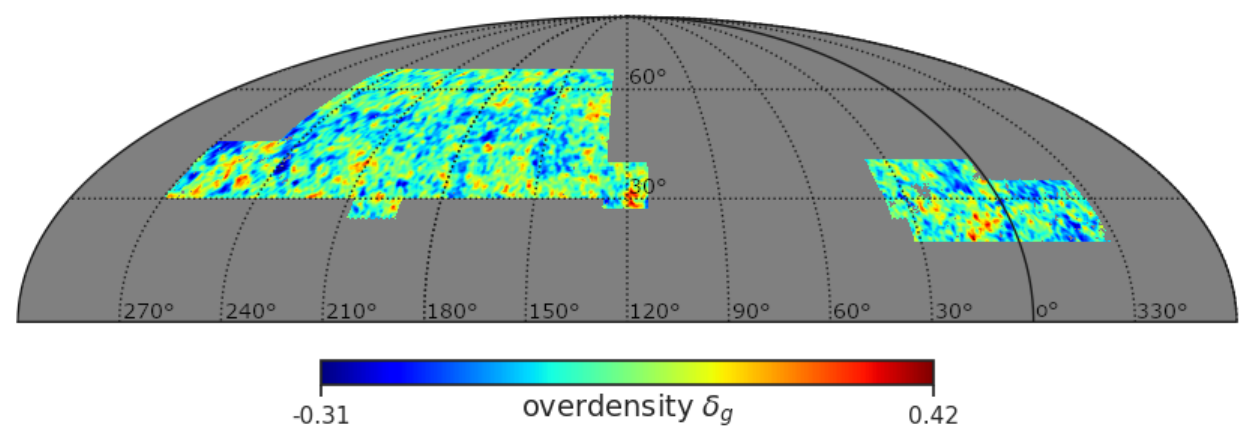}
    \includegraphics[width=0.48\textwidth]{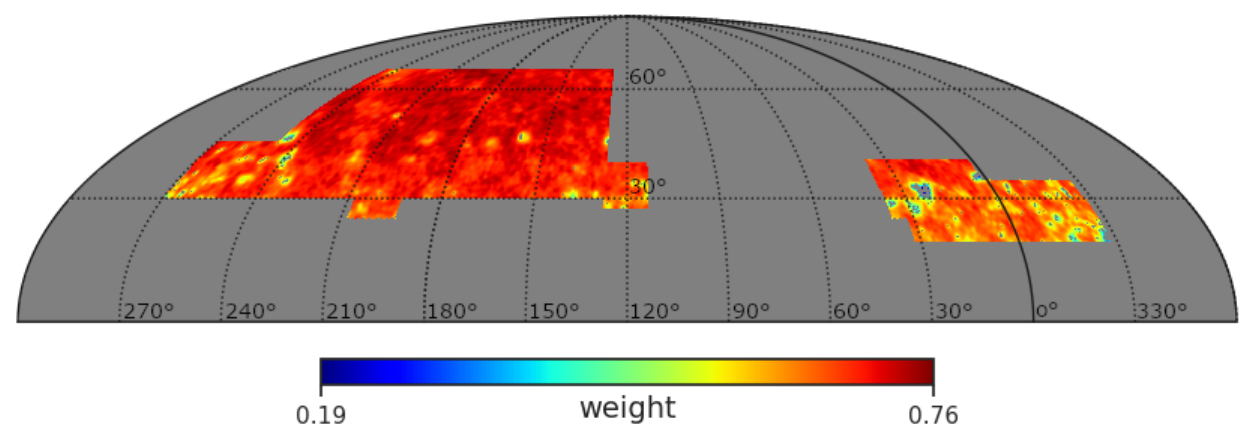}
    \includegraphics[width=0.48\textwidth]{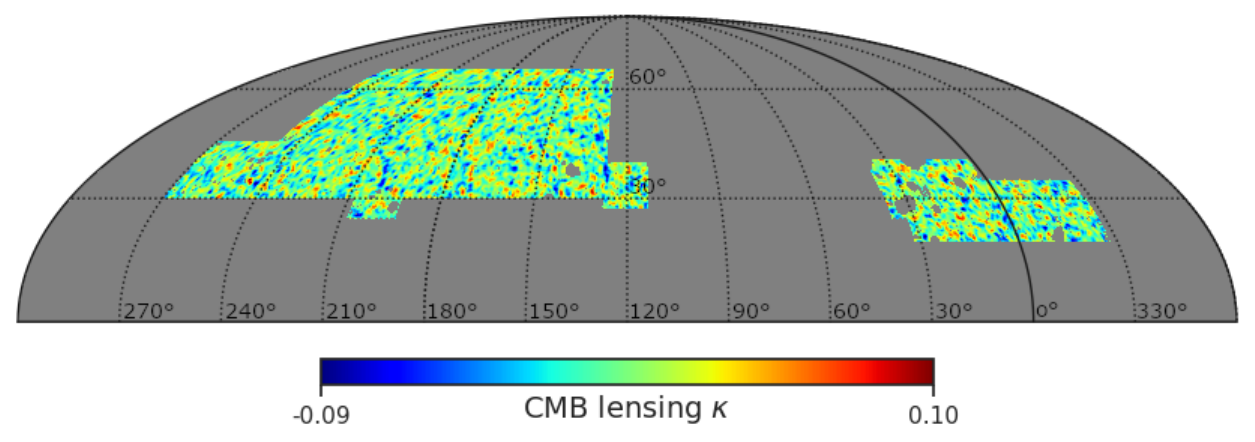}
    \includegraphics[width=0.48\textwidth]{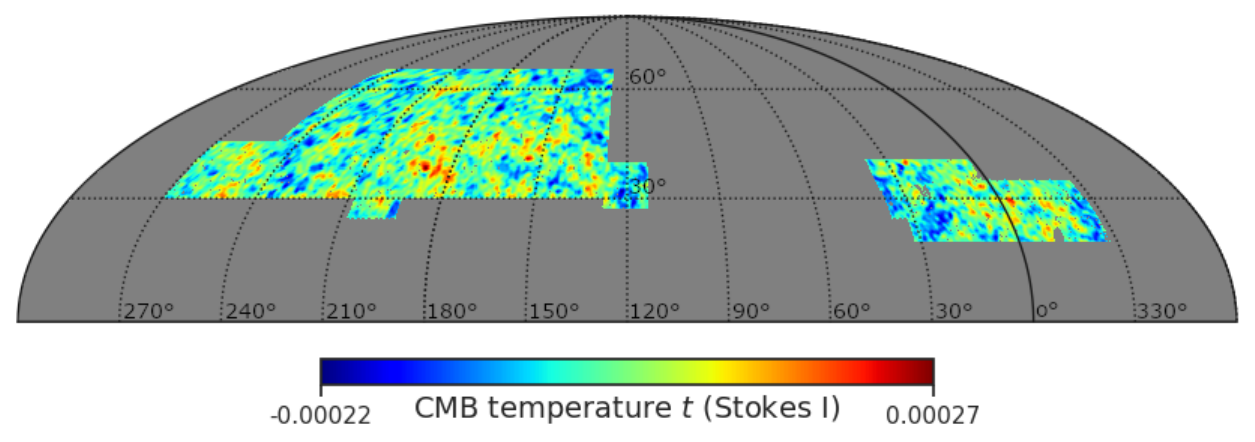}
    \caption{Maps used in this work, smoothed over a scale of $1^{\circ}$, strictly for a visualisation purpose. \textit{Top left}: LoTSS DR2 overdensity, \textit{top right}: LoTSS DR2 completeness based on randoms, \textit{bottom left}: \planck CMB lensing convergence, \textit{bottom right}: \planck CMB temperature. The overdensity and completeness maps include the LoTSS DR2 mask \citep{Hale:2023}, while CMB maps include this and an appropriate mask from the \planck survey.}
    \label{fig:maps}
\end{figure*}

The LOw-Frequency ARray (LOFAR) Two-metre Sky Survey (LoTSS) second data release (DR2) \citep{Shimwell:2022} covers 27\% of the northern sky at 120-168 MHz. It consists of 841 pointings, split into two regions separated by the Galactic plane, spanning 4178 and 1457 square degrees, respectively, and shown in Figure \ref{fig:maps} (top-left). Data reduction was performed using both direction-dependent and independent calibration pipelines, and the source catalogue was created with the source finder PyBDSF \citep{Mohan:2015}. The catalogue derived from the total intensity (Stokes I) maps contains 4,396,228 radio sources.

The completeness and spatial homogeneity of the sample have a complex dependence on flux morphology and signal-to-noise ratio (S/N), defined as the ratio of peak flux density per beam and root mean square noise per beam. The fiducial sample used in this work comprises 1{,}136{,}219 galaxies with a flux density brighter than $1.5 \,{\rm mJy}$, and detected with ${\rm S/N}\geq7.5$, which \citetalias{Hale:2023} finds as the best balance between the number of sources and variation in data compared to random samples. To test the robustness of the cosmological constraints depending on the choice of the sample, we will also present results for an alternative selection corresponding to galaxies above $2.0 \,{\rm mJy}$ with ${\rm S/N}\geq5$ (958{,}438 objects). These cuts are similar to those used in the analysis of the two-point correlation function \citep{Siewert:2020} and the cross correlation with the CMB lensing of LoTSS DR1 \citep{Alonso:2020}.

As described in \citetalias{Hale:2023}, a spatial mask was created by removing regions with tiles that have not been mosaiced together, or have a large number of gaps due to problems in the reduction process, which would lead to strong spatial variations in the flux scale. These regions are mostly located by the outer edges. The resulting unmasked footprint covers 4357 ${\rm deg}^2$, corresponding to a sky fraction $f_{\rm sky}=0.11$. After data cuts and masking, 896{,}637 objects remain in our fiducial catalogue ($1.5\,{\rm mJy}$ flux density cut, ${\rm S/N}\geq7.5$), and 742{,}692 in the $2.0\,{\rm mJy}$, ${\rm S/N}\geq5.0$ sample. We note that because we use pixel coordinates to mask the resulting maps, the final number of objects is slightly different than in \citetalias{Hale:2023} which uses the object coordinates to mask the catalogue.

\subsection{Planck}\label{ssec:data.planck}
We use CMB data released as part of the 2018 \planck analysis \citep{Planck:2020:overview}. We use the `minimum variance' (MV) CMB lensing convergence harmonic coefficients released in \cite{Planck:2020:lensing}, together with the associated sky mask. The harmonic coefficients are transformed into a HEALPix \citep{Gorski:2005} map with resolution parameter $N_{\rm side}$, after truncating them to $(\ell,m)<3\,N_{\rm side}$. In our analysis, we will use a common resolution $N_{\rm side}=512$, corresponding to a pixel size of $\sim6.9\,{\rm arcmin}$. We repeated our analysis using $N_{\rm side}=256$, finding compatible results. The lensing map covers a sky fraction $f_{\rm sky}\simeq0.67$, and overlaps with LoTSS over the whole footprint of the latter.

For the ISW analysis described in Section \ref{ssec:results.isw}, we use the foreground-cleaned temperature fluctuation map produced through the SMICA component separation method, described in \cite{Planck:2020:extraction}, as well as its associated sky mask. Both the mask and map were downgraded to the common resolution $N_{\rm side}=512$.

\section{Methodology}\label{sec:method}
\subsection{Maps}\label{ssec:method.maps}

Figure \ref{fig:maps} shows the maps used in this analysis. In the figure, we smooth them for visualisation purposes with a Gaussian filter with a full width at half-maximum of $1^{\circ}$. The LOFAR maps are limited to the region corresponding to the LOFAR mask, while \planck maps are shown at the intersection between the LOFAR mask and the corresponding CMB mask. Lower and higher resolution falls outside of the scales of interest for this analysis, $50 < \ell < 800$ (see Section \ref{ssec:results.bias}).

We use catalogues of randomly generated sources to account for the spatially varying survey depth. The object positions in the random catalogue should be uncorrelated, while tracing the detection rate, which is not uniform across the footprint. The simulated sources are based on a modified SKA Design Studies Simulated Skies \citep[SKADS, ][]{Wilman:2008, Wilman:2010}, to account for an underestimated number of SFGs at the faintest flux densities \citep[e.g. ][]{Hale:2023a}. The catalogue provides multiple observable properties for simulated sources, which were used in combination with simulations from \citet{Shimwell:2022} to account for the effects of smearing, variation of sensitivity due to elevation or declination, location within the mosaic, proximity to bright sources or edge of the observed field, where the number of mosaiced pointings is smaller. As we do not split the sources between their type (AGN/SFG), redshift, or luminosity, it is the input flux density distribution of the randoms which is the most important, and the modified SKADS represents it well for 144 MHz sources, to below the source detection limit of the survey. The process of generating the random catalogues is described in detail in \citetalias{Hale:2023}. It provides us with simulated `output' number count maps, where output represents galaxies as if they were detected by LOFAR in an idealised case. We use this output map to correct for depth fluctuations, and as a weight for the resulting galaxy overdensity map when computing power spectra (i.e. $w_g(\nv)$ constitutes the mask of the galaxy overdensity map). The overdensity $\delta_g$ is computed as
\begin{equation}
  \delta_g(\nv) = \frac{N_g(\nv)}{\bar{N}_gw_g(\nv)} - 1,
\end{equation}
where $N_g(\nv)$ is the number of galaxies in the pixel lying in the direction of $\nv$. $\bar{N}_g$ is the mean number of objects per pixel, and is estimated as $\bar{N}_g = \langle N_g(\nv) \rangle_{\nv} / \langle w_g(\nv) \rangle_{\nv}$, where $\langle\cdots\rangle_{\nv}$ represents a mean over all pixels within the mask.

\subsection{Redshift distribution}\label{ssec:method.redshift}
An important ingredient of the analysis is the redshift distribution, $p(z)$, of the LoTSS sources, necessary to recover the three-dimensional clustering parameters, which can then be compared with theoretical predictions (eq.~\ref{eq:W_g}). For radio continuum objects, the individual redshifts are not known and cannot be estimated from radio fluxes. At present, we do not have optical identifications and photometric redshift estimates for most of the LoTSS DR2 sources. Therefore, we need to model the underlying $p(z)$ in a more indirect way.

\begin{figure}
    \resizebox{\hsize}{!}{\includegraphics{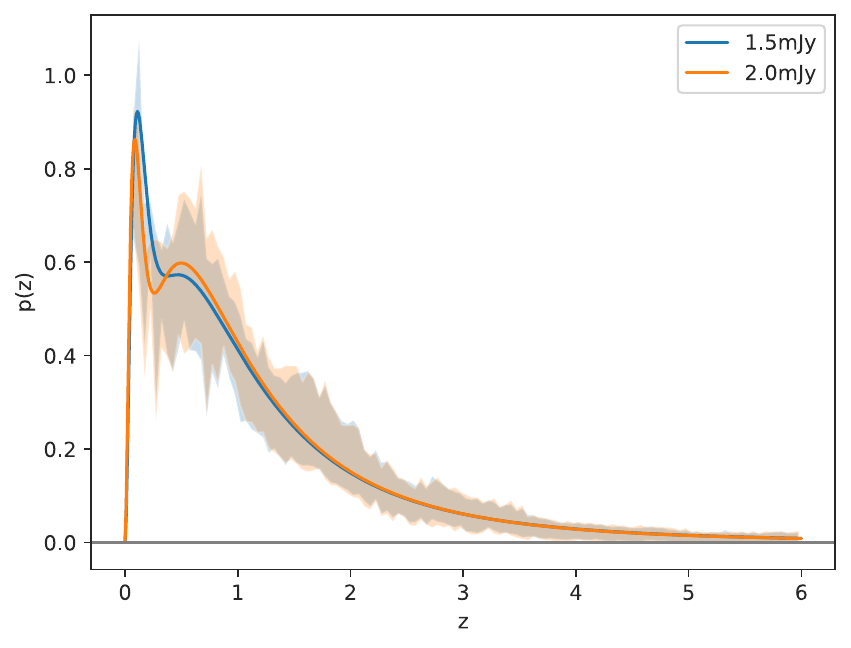}}
    \caption{Redshift distribution based on the three deep fields located within the LoTSS DR2 footprint, for $2\,{\rm mJy}$ and $1.5\,{\rm mJy}$ flux cuts. The thick lines show the models fitted with Eq. \ref{eq:redshift}, and the shaded areas are a $1\sigma$ region from the deep fields measurements. The redshift distribution is limited to $z < 6$.}
    \label{fig:redshift_fit}
\end{figure}
Extragalactic radio sources consist mostly of SFGs and AGNs, although their fractions vary with both redshift and flux density \citep[see][]{Best:2023}. However, limitations in the multi-wavelength coverage of the sample may lead to some uncertainty in the redshifts and classification of sources. In order to calibrate the redshift distribution of our sample, we make use of the LOFAR deep fields observations \citep{Tasse:2021, Sabater:2021}. The Deep Field data consist of three fields: Boötes, ELAIS and Lockman Hole. For each field, a smaller region was defined for which there exists deep multi-wavelength information, of an area equal to 8.6 deg$^2$ in the Boötes field, 6.7 deg$^2$ in ELAIS and 10.3 deg$^2$ in the Lockman Hole field \citep{Kondapally:2021, Duncan:2021}. A redshift and its probability density function were associated with each source using a hybrid method that combined template fitting and machine learning (further details can be found in \citealt{Duncan:2021}). The photometric redshift quality is characterised by normalised median absolute deviation ($\sigma_{\rm{NMAD}}$) ranging from 1.6 to 2\% for galaxies and 6.4 to 7\% for AGNs, while the outlier fraction ($|z_{\rm{phot}} - z_{\rm{spec}}| / (1 + z_{\rm{spec}}) > 0.15$) equals around 2\% for galaxies and 20\% for AGNs. It is worth noting that $\sim5\%$ of the sources satisfying our sample cuts in the deep fields do not have an optical cross-match.

We estimate the redshift distribution for each flux density cut catalogue using a technique based on sampling redshift values from the probability distributions of photometric redshifts, using spectroscopic redshifts where available. Given the full probability distribution over a redshift range for each photometric redshift measurement, we sample a single redshift value over this probability for each object, and build a histogram of such a distribution, binning in $\Delta z = 0.05$. For objects with spectroscopic redshifts available, we always take the reported value (i.e.\ equivalent to zero photo-$z$ uncertainty). We repeat this procedure of histogram creation for each deep field separately, and the number of histograms created for each field is proportional to the number of objects in each field, which makes fields with more observations more significant in the final estimate. We find that the final results do not change after sampling at least 200 histograms in total. The final distribution and its statistical uncertainty is given by the mean and standard deviation calculated over all histogram realisations, and then normalised to a unit integral over the redshift range $0 < z < 6$. This approach to redshift distribution is also described in \citetalias{Hale:2023}. The method is able to combine both photometric and spectroscopic redshifts, and ensures a reasonable estimate of the final uncertainty in the redshift distribution. The uncertainty estimated with this method accounts  both for errors in every single measurement of the photometric redshift, and for differences in redshift distributions between the three deep fields. We found that the errors estimated with this method are significantly larger in comparison to bootstrap sampling over the probability distributions of photometric redshifts, where single redshift distributions are calculated as a sum of probability distributions within the bootstrap samples, and uncertainty is taken as a standard deviation within those.

\begin{table}
    \caption{Constraints on the parameters of the redshift distribution, as given in the equation \ref{eq:redshift}, for the $1.5\,{\rm mJy}$ and $2.0\,{\rm mJy}$ samples.}
    \centering
    \begin{tabular}{l c c c}
    \toprule
    Sample & $z_0$ & $r$ & $a$ \\
    \midrule
    $1.5 \,{\rm mJy}$ & $\zfid$ & $\rfid$ & $\afid$ \\
    $2.0 \,{\rm mJy}$ & $\ztest$ & $\rtest$ & $\atest$ \\
    \bottomrule
    \end{tabular}
    \label{table:redshift}
\end{table}

We model the resulting redshift distribution using a functional form
\begin{equation}\label{eq:redshift}
    p(z) \propto \frac{z^2}{1 + z}\left(\exp\left({\frac{-z}{z_0}}\right) + \frac{r^2}{(1 + z)^a}\right),
\end{equation}
normalised to a unit integral over the redshift range $0 < z < 6$, with $\{z_0, r, a\}$ being free parameters. This form is motivated by the fact that the LoTSS radio sources contain two main populations of objects, AGNs, and SFGs. At low redshifts, we expect their numbers to grow proportionally to the volume for both populations, which motivates the factor of $z^2$, which would be exact for any redshift in a de Sitter model. The factor $1/(1+z)$ provides a simple correction for a $\Lambda$CDM model, and gives a good approximation up to the redshift of $\sim 0.2$. For higher redshifts, the flux density limitation of the sample becomes the dominant aspect, and the form of the luminosity function for each population starts to be important. The AGN radio luminosity function is typically approximated by a double power law, which motivates the power law term, while the SFG radio luminosity function is typically modelled as a Schechter function \citep{2017MNRAS.469.1912B}, which exhibits an exponential cut-off, and motivates the first term. The relative fraction of both contributions is controlled by the parameter $r$. We verified that this three-parameter model provides a good semi-empirical fit and is superior to other simple parameterisations that have been tested. Table \ref{table:redshift} shows the constrained parameters, based on the uncertainties mentioned above, for the $1.5\,{\rm mJy}$ and $2.0\,{\rm mJy}$ samples, while figure \ref{fig:redshift_fit} shows the resulting redshift distributions. The blue and orange bands show the $1\sigma$ constraints measured from the deep fields for the $1.5\,{\rm mJy}$ and $2\,{\rm mJy}$ cuts, respectively, with the corresponding solid lines showing the best-fit model of Eq.~\eqref{eq:redshift} in each case.

\subsection{Bias models}\label{ssec:method.bias}
Given the wide range of redshifts covered by the samples studied, the evolution of the linear galaxy bias over that range must be taken into account. This is non-trivial, as the sample includes several types of extragalactic sources (SFGs and AGNs, in the simplest description), and their relative abundances and intrinsic galaxy biases evolve with $z$. To assess the impact of our assumptions regarding the evolution of the effective bias of the sample, we will consider three different models \citep{Nusser:2015,Alonso:2020}:
\begin{itemize}
    \item A \emph{constant bias} model $b_g(z)=b_g$ represents the simplest case. Although likely an unrealistic model, the corresponding value of $b_g$ can be interpreted as the effective bias of the sample accounting for redshift evolution.
    \item A \emph{constant amplitude} model, in which the bias evolves inversely with the linear growth factor $D(z)$
    \begin{equation}
    b_g(z) = b_{g,D}/D(z).
    \end{equation}
    This model has the advantage of reproducing the expected rise in $b_g(z)$ at high $z$ for a flux-limited sample (assuming a monotonic mass-luminosity relation), while preserving the simplicity of the constant-bias model, with only a single free parameter. In this model, the amplitude of $\Delta_g$ does not change over time at linear order (since $\Delta_m\propto D(z)$). This would correspond to a galaxy distribution that is fixed at some early time and preserves its large-scale properties unchanged \citep{Bardeen:1986,Mo:1996,Tegmark:1998,Coil:2004}.
    \item The two previous models fix the redshift evolution of $b_g(z)$, allowing only its overall amplitude to vary. As a more flexible alternative, we will also use a \emph{quadratic bias} model, in which 
    \begin{equation}
      b_g(z) = b_0 + b_1 z + b_2 z^2,
    \end{equation}
    with $\{b_0,b_1,b_2\}$ free parameters.
\end{itemize}

\subsection{Power spectra}\label{ssec:method.power_spectra}

We use {\tt NaMaster} \citep{Alonso:2019} to compute the angular power spectra of fields defined on a limited region of the sphere using the pseudo-$C_\ell$ estimator \citep{Peebles:1973, Hivon:2002}. We calculate the shot-noise contribution to the galaxy auto-correlation before inverting the pseudo-$C_\ell$ mode-coupling matrix, as \citep{Nicola:2020}
\begin{equation}
  \tilde{N}^{gg}_\ell = \frac{\langle w_g\rangle}{\bar{N}_\Omega},
\end{equation}
where $\bar{N}_\Omega$ is the mean angular number density of galaxies (in units of ${\rm sr}^{-1}$), and $\langle w_g\rangle$ is the value of the mask averaged across the sky. There are reasons to expect departures from a purely Poisson shot noise contribution to $N^{gg}_\ell$. Prominently, in the case of radio surveys, a fraction of sources may have multi-component detections, which effectively leads to a higher shot-noise amplitude than predicted by Poisson statistics \citep{2004MNRAS.351..923B,Tiwari:2022}. Additionally, stochastic and non-local effects in galaxy formation, as well as effects such as halo exclusion, can lead to similar departures from Poissonian shot noise \citep{2016MNRAS.456.3985B,2022MNRAS.514.2198K}. To account for these effects, we marginalise over a free shot noise amplitude $A_{sn}$, which in practice makes the pipeline sensitive only to non-flat contributions to the galaxy auto-correlation.

To calculate the statistical uncertainties of our measurements of $C_\ell^{xy}$, we use a jackknife resampling procedure \citep{Norberg:2009}. We divide the LoTSS DR2 footprint into 54 similarly sized rectangular areas. We find this number of regions to provide a good balance between the small-scale and large-scale errors. Then, removing one of these areas at a time, we calculate the power spectra in the resulting footprint. The power spectrum covariance is then calculated as
\begin{equation}
  {\sf Cov}(C_\ell^x, C_\ell^y) = \frac{N_{\rm JK} - 1}{N_{\rm JK}} \sum_{i=1}^{N_{\rm JK}}(C_\ell^{x,i} - \bar{C}_\ell^x)(C_\ell^{y,i} - \bar{C}_\ell^y),
\end{equation}
where $x$ and $y$ stand for $(gg,g\kappa,gT)$, $N_{\rm JK}$ is the number of jackknife samples, $C_\ell^{x,i}$ is the power spectrum measured in the $i$-th sample, and $\bar{C}_\ell^x\equiv\sum_i C_\ell^{x,i}/N_{\rm JK}$ is the average over jackknife samples. To validate this estimate of the covariance matrix, we compared it with the analytical prediction assuming all fields studied can be described by Gaussian statistics \citep[e.g.][]{Garcia-Garcia:2019}. Both estimates were found to be in good agreement. We also report a correlation matrix
\begin{equation}
\label{eq:correlation}
    {\sf r}_{ij} = {\sf Cov}_{ij}\,/\sqrt{{\sf Cov}_{ii}{\sf Cov}_{jj}},
\end{equation}
where ${\sf r}$ is the correlation coefficient, while $i$ and $j$ are corresponding indices of the covariance matrix.

\subsection{Likelihood inference}\label{ssec:method.like}
We assume the power spectrum data follow a Gaussian distribution, and we estimate the log-likelihood as
\begin{equation}
    \chi^2 \equiv -2\log p({\bf d}|{\bf q}) = [{\bf d}-{\bf t}({\bf q})]^T{\sf Cov}^{-1}[({\bf d}-{\bf t}({\bf q})],
\end{equation}
where ${\bf d}$ denotes the data vector, consisting of combinations of $C_\ell^{gg}$ and $C_\ell^{g\kappa}$, as well as the deep field measurements of the redshift distribution described in Section \ref{ssec:method.redshift}. We do not use $C_\ell^{gT}$ as part of the inference scheme, as we only report its significance. ${\bf t}({\bf q})$ is the theoretical prediction for ${\bf d}$ given a set of parameters ${\bf q}$, describing both the power spectra and the redshift distribution (as parameterised in Eq.~\ref{eq:redshift}). The covariance matrix ${\sf Cov}$ incorporates the correlated uncertainties of the different elements of ${\bf d}$. We assume that the power spectrum and redshift distribution measurements are uncorrelated, while retaining all potential correlations between different power spectra (at different scales and for different fields). In other words, the covariance matrix elements for $(p(z),C_\ell^{xy})$ are set to zero. The covariance of the measured redshift distribution was assumed to be diagonal, with errors estimated via sampling as described in Section \ref{ssec:method.redshift}.

We report the significance of the $C^{g\kappa}_\ell$ power spectrum and the ISW signal as the square root of the difference in $\chi^2$ between a null hypothesis, defined as $C_\ell=0$, and the best-fit model ($\sqrt{\Delta\chi^2}$). We calculate the reduced chi-squared as:
\begin{equation}
  \chi^2_{\nu} = \frac{\chi^2}{\nu},
\end{equation}
where $\nu$ stands for the number of degrees of freedom equal to the number of observations minus the number of fitted parameters. The number of observations includes the data points from the  $C_{\ell}^{gg}$ and $C_{\ell}^{g\kappa}$, while the number of fitted parameters includes the bias parameters, amplitude of shot noise, and $\sigma_8$. Data points from deep fields and parameters from the redshift distribution modelling are not included while reporting those statistics. Finally, we report the `probability to exceed' (PTE), calculated in terms of the $\chi^2$ as:
\begin{equation}
  \mathrm{PTE}(\chi^2, \nu) = 1 - F(\chi^2, \nu),
\end{equation}
where $F$ denotes the $\chi^2$ cumulative distribution function.

\begin{table}
    \caption{List of all parameters used in the inference with the corresponding priors and initial values. We require $b_0$ and $b_1$ to be positive, in order to obtain positive $a$, and to be increasing at low redshifts. We allow $b_2$ to be negative, as it can both increase and decrease the bias evolution. The initial values are drawn uniformly from a range centred at the `middle point' and bounded by plus and minus 20\% of this value.}
    \centering
    \begin{tabular}{l l c c}
    \toprule
    & & prior & middle point \\
    \midrule
    constant bias & $b_g$ & positive & 2.0 \\
    const. amplitude bias & $b_{g,D}$ & positive & 1.5 \\
    quadratic bias & $b_0$ & positive & 1.5 \\
    & $b_1$ & positive & 1.0 \\
    & $b_2$ & none & 0.1 \\
    redshift distribution & $z_{0}$ & positive & 0.05 \\
    & $a$ & none & 5.0 \\
    & $r$ & positive & 0.2 \\
    shot noise amplitude & $A_{sn}$ & [0.8, 1.4] & 1.1 \\
    matter fluctuations & $\sigma_8$ & positive & 0.81 \\
    \bottomrule
    \end{tabular}
    \label{table:parameters}
\end{table}
 
To explore the posterior distribution function we make use of rejection sampling via Monte-Carlo Markov Chains (MCMC) as implemented in the public {\tt emcee} code \citep{emcee:2013}. Table \ref{table:parameters} lists the parameters of interest explored in Section \ref{sec:results} together with their priors. We free up the value of $\sigma_8$ only in Section \ref{ssec:results.cosmology}. The MCMC chains were generated using 32 walkers and a convergence condition ensuring that the number of samples is equal to or higher than 40 times the mean of the auto-correlation scale for all the inferred parameters.

\section{Results}\label{sec:results}

\subsection{Power spectra}\label{ssec:results.power-spectra}

\begin{figure*}
    \centering
    \subfloat{
    \includegraphics[width=8.5cm]{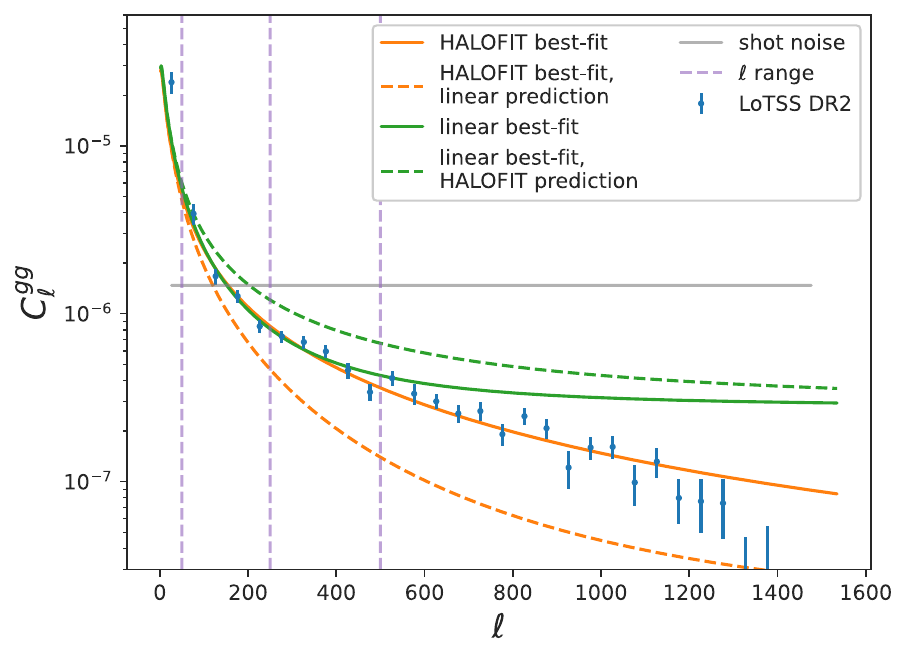}
    }
    \subfloat{
    \includegraphics[width=8.5cm]{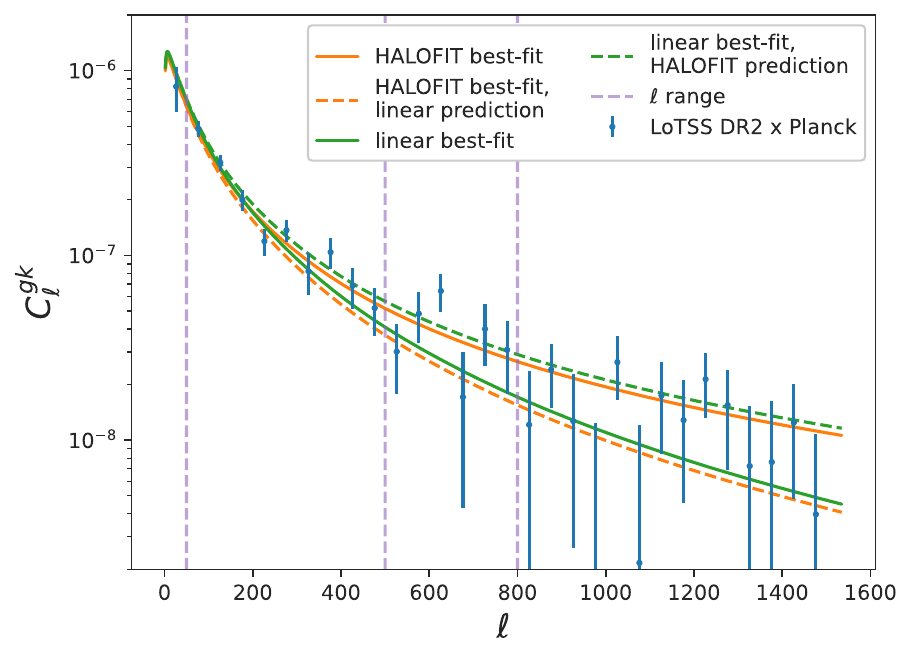}
    }
    \caption{Comparison of the linear and {\tt HALOFIT} matter power spectrum for the auto- and cross-correlation (\textit{left} and \textit{right}, respectively). We note that the shot noise is reported for each multipole separately, while the correlation signal is calculated in the bins of 50 multipoles. The solid lines show the best-fit results with different 3D power spectrum models, while dashed lines show the models with the same resulting best-fit parameters as obtained for solid lines, but with only matter power spectrum changed to the other model. Hence, the difference between the corresponding solid and dashed lines stems only from a difference between linear and {\tt HALOFIT} models. The vertical dashed lines mark the multipole ranges used in this analysis: the fiducial $50 \leq \ell \leq 250$ and $50 \leq \ell \leq 500$, as well as larger $50 \leq \ell \leq 500$ and $50 \leq \ell \leq 800$, for $C_\ell^{gg}$ and $C_\ell^{g\kappa}$ respectively. The fits shown here were made on the fiducial multipole range, where differences between the linear and {\tt HALOFIT} models are between $1\sigma-2\sigma$ of errors on data measurements.}
    \label{fig:linear_vs_halofit}
\end{figure*}

\begin{figure}
    \resizebox{0.9\hsize}{!}{\includegraphics{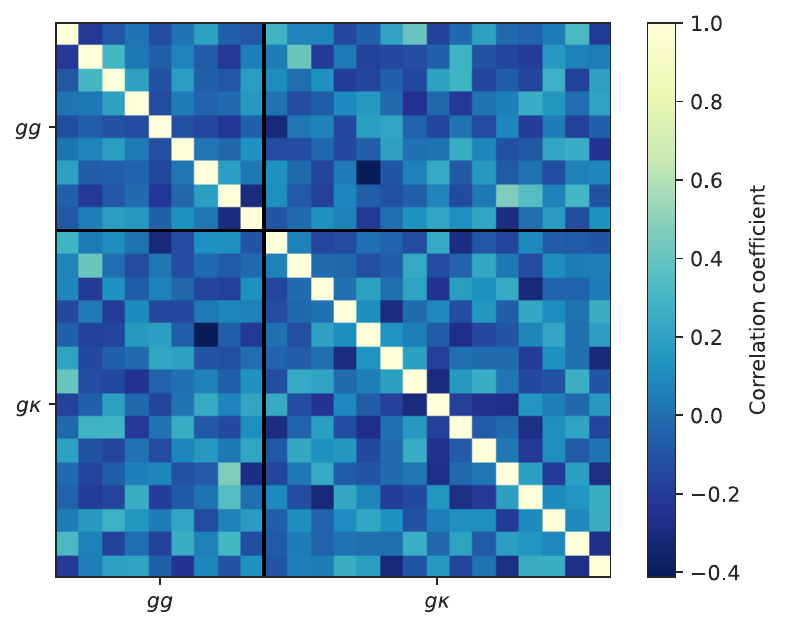}}
    \caption{Correlation matrix (${\sf r}_{ij}$, see eq. \ref{eq:correlation}) for $C_\ell^{gg}$, $C_\ell^{g\kappa}$ power spectra. Multipole ranges $50 \leq \ell \leq 500$ and $50 \leq \ell \leq 800$ are shown, which are the largest ranges used in this work, in bins of $\Delta \ell = 50$.}
    \label{fig:correlation_matrix}
\end{figure}

Figure \ref{fig:linear_vs_halofit} shows the measurements of the LoTSS DR2 auto-spectrum, and its cross-correlation with the \planck lensing map (left and right panels respectively). The solid grey line in the left panel shows the expected contribution from shot noise. As expected, given the broad redshift range covered by the sample, the auto-correlation has a featureless, roughly power-law-like behaviour, which is detected at relatively high significance over all the scales explored. The cross-correlation is also clearly detected to scales $\ell\sim800$. Quantifying the significance of this detection as described in Section \ref{ssec:method.like} (from the $\chi^2$ difference between best-fit model and null hypothesis), including these scales, we obtain a signal-to-noise  of:
\begin{equation}
  \left(\frac{S}{N}\right)_{\ell\leq800}=\sigmaCgkhigh.
\end{equation}
This is one of the most significant detections of the cross-correlation between radio galaxies and CMB lensing so far, comparable to the significance of the correlation with the NVSS sample over a much larger area \citep{Planck:2013mth}. Considering only the fiducial scales $\ell\leq500$ that we will include in the analysis, the significance is
\begin{equation}
  \left(\frac{S}{N}\right)_{\ell\leq500}=\sigmaCgklow.
\end{equation}
This is a factor $\sim3.6$ higher than the detection in \citetalias{Alonso:2020} using LoTSS DR1, in good agreement with the expectation given the relative increase in area between both releases (assuming that $S/N$ scales as $\sqrt{f_{\rm sky}}$). For the auto-correlation, we obtain the signal-to-noise ratio of $\sigmaCgghigh\sigma$ at $\ell \leq 500$, and $\sigmaCgglow\sigma$ at $\ell \leq 250$.

Figure \ref{fig:linear_vs_halofit} also shows the best-fit predictions for both power spectra using the constant amplitude bias model and linear (green line) and {\tt HALOFIT} (orange line) predictions, as well as corresponding models resulting from using the same best-fit parameters as found earlier, but changing only the matter power spectrum (both dashed lines). Comparing the same colour solid and dashed lines, linear and {\tt HALOFIT} predictions begin to differ from one another by more than $2\sigma$ of the statistical uncertainties in our measurements of $C_\ell^{gg}$ at $\ell=250$, and about $1\sigma$ of the error of $C_\ell^{g\kappa}$ at $\ell=500$ in case of the cross-correlation. Using the approximation $\theta \simeq 180^{\circ} / \ell$, those scales translate to $\sim0.72^{\circ}$ and $\sim0.36^{\circ}$, respectively. On the other hand, at the median redshift $z_{\rm med}\simeq0.82$, based on the fitted distribution, these angular scales correspond to wave numbers $k=0.13\,h\,{\rm Mpc}^{-1}$ and $0.26\,h\,{\rm Mpc}^{-1}$ respectively. We can thus use this to define conservative scale cuts for which the linear bias model can be considered. Our fiducial scale cuts will therefore be reported here as $\ell<(250, 500)$. To eliminate any residual systematics associated with large-scale survey depth variations, we will also remove the first bandpower in the galaxy auto-correlation.

\bgroup
\def\arraystretch{1.5}
\begin{table*}
    \caption{Comparison of bias estimates for constant bias and constant amplitude models, using different power spectrum measurements at the fiducial $\ell < (250, 500)$ scale cuts, together with the {\tt HALOFIT} matter power spectrum.}
    \centering
    \begin{tabular}{l c c c c c c c c}
    \toprule
    & \multicolumn{4}{c}{$b_g(z) = b_g$} & \multicolumn{4}{c}{$b_g(z) = b_{g,D} / D(z)$} \\
    {} & $b_g$ & $A_{sn}$ & $\chi^2_{\nu}$ & PTE & $b_{g,D}$ & $A_{sn}$ & $\chi^2_{\nu}$ & PTE \\
    \midrule
    $C_\ell^{gg}$      &  $1.86^{+0.14}_{-0.14}$ &  $0.93^{+0.10}_{-0.08}$ &               1.7 &          19\% &  $1.53^{+0.09}_{-0.11}$ &  $0.93^{+0.11}_{-0.08}$ &             1.7 &        19\% \\
    $C_\ell^{g\kappa}$             &  $2.16^{+0.10}_{-0.09}$ &                         &               1.2 &          30\% &  $1.39^{+0.06}_{-0.06}$ &                         &             1.2 &        32\% \\
    $C_\ell^{gg}$ \& $C_\ell^{g\kappa}$ &  $2.08^{+0.09}_{-0.09}$ &  $0.89^{+0.08}_{-0.06}$ &              1.4 &          18\% &  $1.41^{+0.06}_{-0.05}$ &  $1.01^{+0.08}_{-0.09}$ &            1.2 &        25\% \\
    \bottomrule
    \end{tabular}
    \label{table:correlations}
\end{table*}
\egroup

It is worth noting, however, that using a linear bias model applied to the non-linear matter power spectrum has been empirically found to extend the validity of the model to mildly non-linear scales \citep{Pandey:2020,Sugiyama:2022,Porredon:2022}. In some cases, we will therefore also report results for less conservative scale cuts $\ell<(500,800)$ (corresponding to $k_{\rm max}=0.26\,h\,{\rm Mpc}^{-1}$ and $0.42\,h\,{\rm Mpc}^{-1}$ respectively). We stress, however, that these results should be interpreted with care, since they rely on the validity of the linear bias model over mildly non-linear scales. This could be quantified via numerical simulations including a physics-based model for the galaxy-halo relation for SFGs and AGNs, but this lies beyond the scope of this work. The correlation matrix of the joint $(C_\ell^{gg},C^{g\kappa}_\ell)$ data vector after imposing these scale cuts is shown in Fig. \ref{fig:correlation_matrix}. As evidenced by this plot, the uncertainties between different bandpowers are largely uncorrelated.

\subsection{Constraining bias}\label{ssec:results.bias}
We now use the measurements presented in the previous section to constrain the bias of radio sources in the LoTSS DR2 sample. For now, we will fix all cosmological parameters to the best-fit \planck cosmology \citep{Planck:2020:overview}, and will only vary bias and $p(z)$ parameters.

We begin by comparing our two 1-parameter bias models, the constant-bias and constant-amplitude (or $1/D(z)$) parametrisations, when constrained by different combinations of correlation functions, but in all cases using our fiducial scale cuts $\ell<(250,500)$, and the {\tt HALOFIT} matter power spectrum. Table \ref{table:correlations} shows the constraints on the bias and the shot-noise amplitude $A_{sn}$ obtained using our fiducial scale cuts when including only the galaxy auto-correlation (first row), the cross-correlation (second row), and both (third row). We find that, while both models are able to provide a good fit to the auto- and cross-correlations separately, the $1/D(z)$ model provides a better representation of both signals simultaneously. The combined constraint $\biasd$ is compatible with the individual constraints from $C_\ell^{gg}$ and $C_\ell^{g\kappa}$, which are also in agreement with each other. The constant-bias model, in turn, finds broadly incompatible best-fit values for $b_g$, and the model is a worse fit for the combined data vector than the $1/D(z)$ model. It is also interesting to note that, with the conservative scale cuts applied here, the bias is better constrained with $C_\ell^{g\kappa}$ than with the galaxy auto-spectrum. When including the CMB lensing cross-correlation, there is then significant evidence that the effective bias of the sample grows with redshift, as would be expected for most flux-limited samples.

\bgroup
\def\arraystretch{1.5}
\begin{table*}
    \caption{Comparison of bias estimates for constant amplitude and quadratic models, using different multipole ranges and modelling of matter power spectrum, obtained with both $C_\ell^{gg}$ and $C_\ell^{g\kappa}$. The last column shows the number of data points.}
    \centering
    \begin{tabular}{l l c c c c c c c c c c}
        \toprule
        & & \multicolumn{4}{c}{$b_g(z) = b_{g,D} / D(z)$} & \multicolumn{6}{c}{$b_g(z) = b_0 + b_1 z + b_2 z^2$} \\
        {} & & $b_{g,D}$ & $A_{sn}$ & $\chi^2_{\nu}$ & PTE & $b_0$ & $b_1$ & $b_2$ & $A_{sn}$ & $\chi^2_{\nu}$ & PTE \\
        \midrule
        $\ell < (250, 500)$  & linear  &  $1.54^{+0.06}_{-0.06}$ &  $1.19^{+0.06}_{-0.06}$ &            1.4 &        16\% &  $1.54^{+0.21}_{-0.24}$ &  $0.67^{+0.67}_{-0.46}$ &  $0.19^{+0.25}_{-0.28}$ &  $1.21^{+0.06}_{-0.07}$ &               1.8 &           6.8\% \\
        & {\tt HALOFIT} &  $1.41^{+0.06}_{-0.05}$ &  $1.01^{+0.08}_{-0.09}$ &            1.2 &        25\% &  $1.56^{+0.19}_{-0.21}$ &  $0.57^{+0.50}_{-0.39}$ &  $0.06^{+0.20}_{-0.17}$ &  $0.98^{+0.10}_{-0.09}$ &               1.4 &           17\% \\
        $\ell < (500, 800)$ & linear  &  $1.65^{+0.04}_{-0.04}$ &  $1.14^{+0.02}_{-0.02}$ &            1.8 &       1.6\% &  $1.60^{+0.19}_{-0.26}$ &  $0.83^{+0.77}_{-0.59}$ &  $0.23^{+0.28}_{-0.34}$ &  $1.15^{+0.02}_{-0.02}$ &               1.9 &          0.9\% \\
        & {\tt HALOFIT} &  $1.44^{+0.04}_{-0.04}$ &  $1.04^{+0.02}_{-0.02}$ &            1.5 &        7.8\% &  $1.51^{+0.16}_{-0.18}$ &  $0.65^{+0.53}_{-0.45}$ &  $0.11^{+0.23}_{-0.22}$ &  $1.04^{+0.02}_{-0.02}$ &               1.6 &           3.9\% \\
        \bottomrule
    \end{tabular}
    \label{table:bias}
\end{table*}
\egroup

Having established that the constant bias model is mildly disfavoured, we use constant-amplitude and quadratic bias models to further compare the bias estimates between the linear and {\tt HALOFIT} models of the power spectrum at different scales. The results are shown in Table \ref{table:bias}. The linear and {\tt HALOFIT} models are in broad agreement within $\sim2 \sigma$ at the fiducial scale cuts $\ell < (250, 500)$. At the less conservative scale cuts $\ell < (500, 800)$, the agreement is significantly poorer, and neither model is able to provide a good fit to the data (with PTEs at around 3\% and 13\% for linear and {\tt HALOFIT} models respectively). This shows that the linear bias assumption employed here is not a reliable representation of the data on these mildly non-linear scales. It is worth noting that, for either choice of scale cuts, the linear power spectrum model achieves a consistently poorer $\chi^2$ than {\tt HALOFIT} ($\Delta\chi^2\simeq6$ for $\ell<(500,800)$, with a significantly worse PTE). The linear power spectrum model also generally prefers a $10\%$ to $20\%$ higher value of the shot-noise amplitude $A_{sn}$, to compensate for the lower small-scale power in comparison with {\tt HALOFIT}.

\begin{figure}
    \centering
    \resizebox{\hsize}{!}{\includegraphics{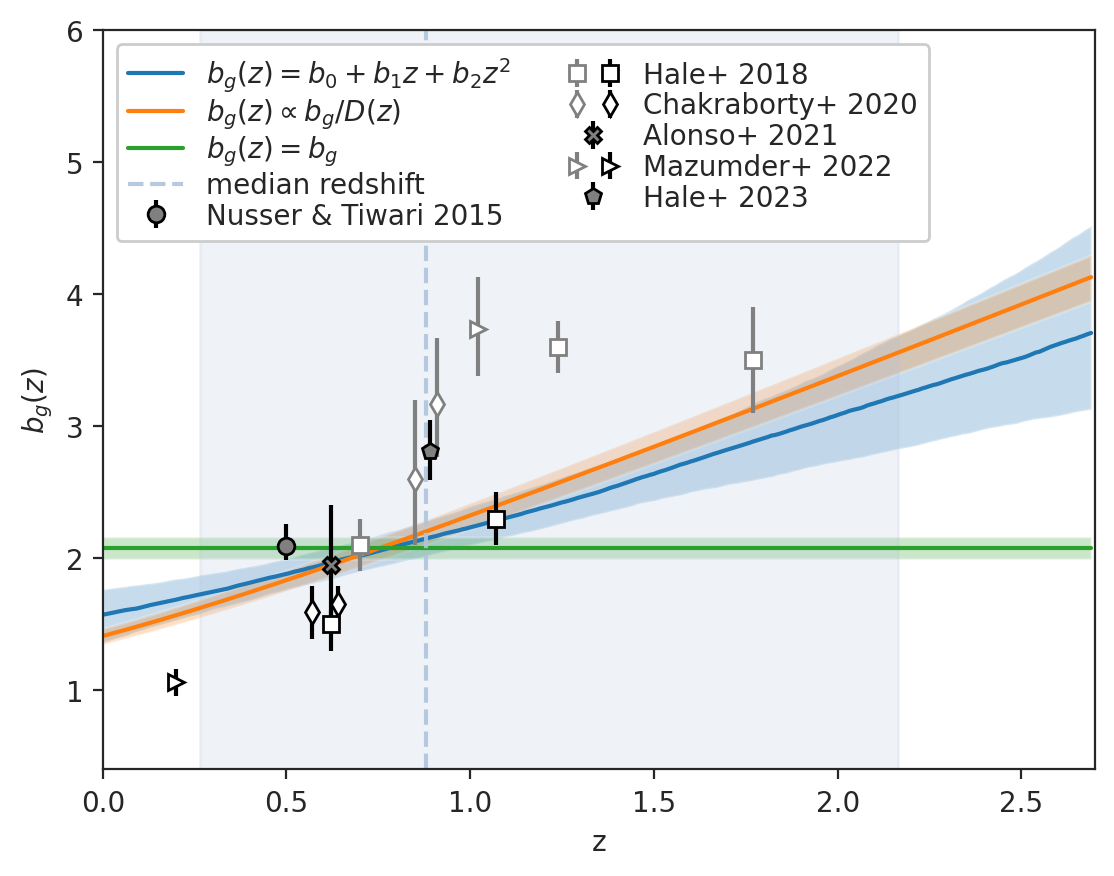}}
    \caption{Bias constraints for three different models based on $C_\ell^{gg}$ and $C_\ell^{g\kappa}$ at the fiducial $\ell < (250, 500)$, and for the {\tt HALOFIT} matter power spectrum. The blue vertical line shows the LoTSS DR2 median redshift and 68 percentiles based on the deep fields $N(z)$. The markers show bias estimates for radio galaxies known from the literature \citep{Nusser:2015, Hale:2018, Alonso:2020, Chakraborty:2020, Mazumder:2022, Hale:2023}. Markers with white fillings and grey/black borders stand for AGNs/SFGs respectively, while markers with grey fillings and black borders denote mixed populations.}
    \label{fig:bias}
\end{figure}

The combination of $C_\ell^{gg}$ and $C_\ell^{g\kappa}$ allows us to successfully constrain the quadratic bias model. The constraints on the bias evolution obtained from the joint data vector, analysed under the {\tt HALOFIT} model, for the three bias evolution models explored here, are shown in Fig. \ref{fig:bias}. The figure also shows other existing estimates of the bias of various radio galaxy samples \citep{Nusser:2015, Hale:2018, Chakraborty:2020, Mazumder:2022}, which are of different depth and different ratios of AGNs and SFGs, as well as the results obtained in \citetalias{Hale:2023} using the galaxy correlation function. The figure also shows, as a vertical shaded band, the median and 68 percentiles of the redshift distribution estimated from the deep fields. Using the median survey redshift, the $1/D(z)$ model predicts a value of the bias $\biasval$. This is in good agreement with the prediction from the other two bias models at the same redshift. Future releases of LoTSS data may allow us to better constrain the values of the quadratic bias model, which imposes fewer assumptions on bias evolution. With the current data, the predictions of the quadratic bias model are in good agreement with those of the $1/D(z)$ model. From these results, we are led to conclude that the more reliable setup to carry out cosmological analyses with the LoTSS data is to adopt the conservative scale cuts $\ell < (250, 500)$ paired with the {\tt HALOFIT} matter power spectrum and the $1/D(z)$ bias model, as the simplest parametrisation able to describe all the data used in the analysis.

The different results from the literature shown in Fig. \ref{fig:bias} are in rough agreement with our measurements. It is worth noting, however, that each of these works was carried out on samples of radio galaxies with different depths and ratios of AGNs and SFGs, also using different bias parametrisations, and hence a direct comparison is not possible. The most direct comparisons can be made with the measurements of \citetalias{Alonso:2020}, using a sample at flux higher than $2\,{\rm mJy}$ and an S/N higher than $5$ defined on the LoTSS DR1 catalogue, and with the estimate of \citetalias{Hale:2023}, based on the angular auto-correlation function of LoTSS DR2 for a flux limit of $1.5\,{\rm mJy}$ and an S/N cut of $7.5$. \citetalias{Alonso:2020} found the joint constraints on galaxy bias to depend strongly on the assumed redshift distribution of the sample, although the cross-correlation alone was extremely robust against this systematic. In this case, for the $1/D(z)$ model, \citetalias{Alonso:2020} finds $b_{g,D}=1.46\pm0.28$, in agreement with our findings ($\biasdtest$ at $2.0\,{\rm mJy}$, at $5\,\rm{S/N}$). \citetalias{Hale:2023} used a setup that is much more similar to ours, although only studying the LoTSS auto-correlation. Using the linear matter power spectrum and a scale range $36 < \ell < 360$ (assuming a conversion $\ell = 180^{\circ} / \theta$ between real and harmonic scales), \citetalias{Hale:2023} found $b_{g,D} = 1.79^{+0.15}_{-0.14}$, which is at $1.3\sigma$ difference from our corresponding setup which yields $b_{g,D} = 1.61 \pm 0.11$, while using only the auto-correlation at $50 < \ell < 250$ and the linear matter power spectrum. Our bias estimate becomes lower, $b_{g,D} = 1.54 \pm 0.06$, and the difference becomes larger, at a level of $1.8\sigma$, if we add the cross-correlation, as shown in the first row of table \ref{table:bias}. The difference becomes even larger, at a level of $2.3\sigma$, if we assume the {\tt HALOFIT} matter power spectrum, which gives better fits in our case and evaluates to $\biasd$, in comparison to $b_{g,D} = 1.75^{+0.16}_{-0.15}$ from \citetalias{Hale:2023}, while also using the {\tt HALOFIT}. We follow upon this in Sect.~\ref{sec:discussion}.

\subsection{Bias evolution and clustering redshifts}\label{ssec:results.tomographer}

\begin{figure}
    \resizebox{\hsize}{!}{\includegraphics{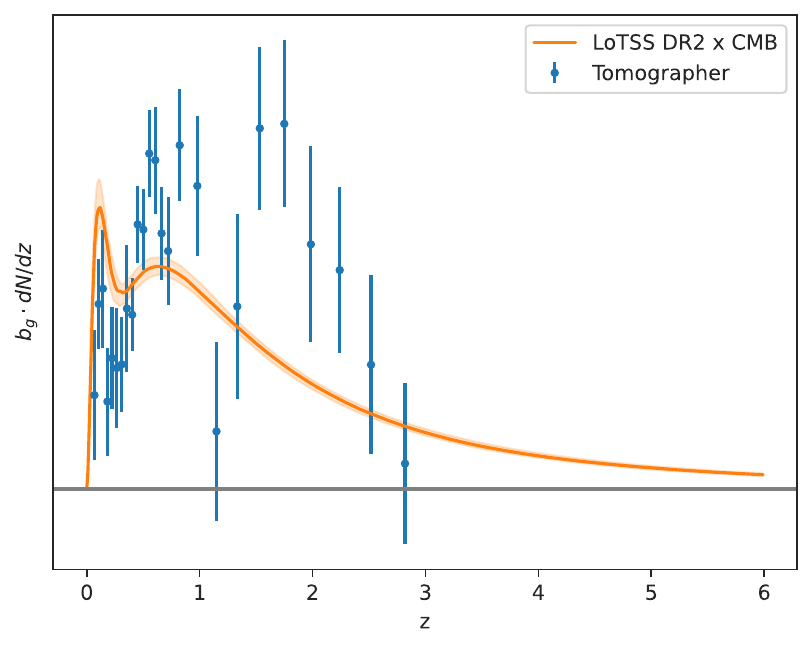}}
    \caption{Comparison of our fit to redshift distribution and bias from our fiducial approach and $b_{g,D}/D(z)$ bias modelling, against the results from Tomographer.}
    \label{fig:tomographer}
\end{figure}

Since all the bias models explored here assume some form of bias evolution, we carry out one additional test of their validity, making use of the clustering redshifts technique \citep{Newman:2008, Menard:2013, Scottez:2016}. Clustering-based redshift estimation uses a set of angular cross-correlations between a sample for which the redshift distribution is unknown and a reference spectroscopic sample with known redshifts, to infer the unknown redshift distribution. Due to the degeneracy between the redshift distribution and the galaxy bias of the target sample in setting the amplitude of the cross-correlations, the technique is in fact only able to constrain the combination $b_g(z)\,p(z)$. Although this degeneracy with bias evolution is one of the drawbacks of the clustering redshifts technique, we can use it to our advantage in order to validate our assumptions regarding bias evolution. We estimate $b_g(z)\,p(z)$ for our sample using the public tool {\tt Tomographer}\footnote{\url{http://tomographer.org}.}\citep{Chiang:2019}, which uses about 2 million spectroscopic objects covering about 10,000 square degrees, based on samples of galaxies and quasars from the Sloan Digital Sky Survey \citep[SDSS, ][]{Strauss:2002, Blanton:2005, Schneider:2010, Reid:2016, Paris:2017, Ata:2018, Bautista:2018}. We compare the result from {\tt Tomographer} with the product of the redshift distribution obtained from the deep fields and the best-fit $1/D(z)$ bias model. Figure \ref{fig:tomographer} shows the results, with {\tt Tomographer} measurements shown as points with error bars, and the $68\%$ confidence interval of our estimated $b_g(z)\,p(z)$ shown as an orange band. Both results are in broad agreement, but there is some potential evidence of a higher bias at $z\gtrsim1.5$, which could be confirmed with future LoTSS data releases, or with a dedicated cross-correlation analysis involving a dense optical galaxy sample at those redshifts \citep[e.g.][]{unwise,StoreyFisher:2023}.

\subsection{Constraining $\sigma_8$}\label{ssec:results.cosmology}

\begin{figure}
    \resizebox{\hsize}{!}{\includegraphics{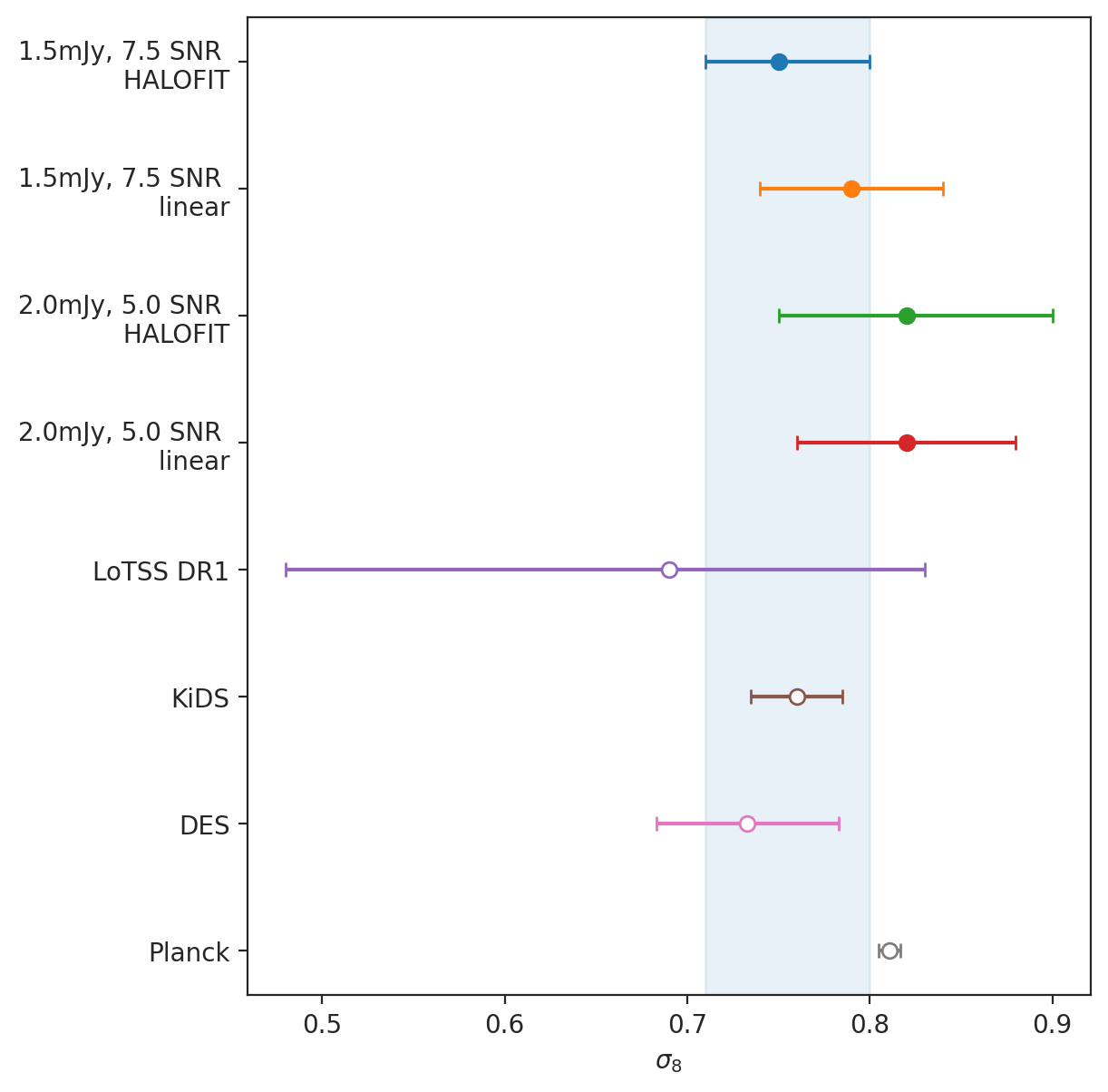}}
    \caption{Constraints on $\sigma_8$ using $C_\ell^{gg}$ and $C_\ell^{g\kappa}$ at the fiducial $\ell < (250, 500)$, the $b_{g,D}/D(z)$ bias modelling, {\tt HALOFIT} matter power spectrum, and \planck cosmology assumed for parameters other than $\sigma_8$. The top bar shows constraints form the LoTSS DR1 \citep{Alonso:2020}, and the three bottom bars present \planck \citep{Planck:2020:overview}, KiDS \citep{Heymans:2021}, and DES \citep{Abbott:2022}.}
    \label{fig:sigma_8}
\end{figure}

\bgroup
\def\arraystretch{1.5}
\begin{table*}
    \caption{Comparison of $\sigma_8$ estimates at two different choices of data cuts, using both $C_\ell^{gg}$ and $C_\ell^{g\kappa}$ and the fiducial scale cut $\ell < (250, 500)$ and {\tt HALOFIT} matter power spectrum. The number of data points in the correlation functions is 13.}
    \centering
    \begin{tabular}{l l c c c c c}
    \toprule
    Sample & Matter power spectrum & $\sigma_8$ & $b_{g,D}$ & $A_{sn}$ & $\chi^2$ & PTE\\
    \midrule
    ($1.5 \,{\rm mJy}$, ${\rm S/N} > 7.5$) & {\tt HALOFIT} & $0.75^{+0.05}_{-0.04}$ &  $1.62^{+0.21}_{-0.19}$ &  $0.97^{+0.09}_{-0.09}$ &            1.2 &        26\% \\
    
    & linear & $0.79^{+0.05}_{-0.05}$ &  $1.62^{+0.20}_{-0.18}$ &  $1.17^{+0.08}_{-0.08}$ &            1.5 &        12\% \\
    
    ($2.0 \,{\rm mJy}$, ${\rm S/N} > 5.0$)     & {\tt HALOFIT} & $0.82^{+0.08}_{-0.07}$ &  $1.38^{+0.25}_{-0.22}$ &  $1.17^{+0.10}_{-0.09}$ &            1.4 &        20\% \\
    
    & linear & $0.82^{+0.06}_{-0.06}$ &  $1.49^{+0.20}_{-0.18}$ &  $1.30^{+0.06}_{-0.08}$ &            1.8 &        5.7\% \\
    \bottomrule
    \end{tabular}
    \label{table:sigma_8}
\end{table*}
\egroup

We put constraints on the $\sigma_8$ parameter by using $C_\ell^{gg}$ and $C_\ell^{g\kappa}$ at the fiducial $\ell < (250, 500)$ scale cuts, together with the deep fields $p(z)$, the {\tt HALOFIT} matter power spectrum, and the $1/D(z)$ bias model, as justified in previous sections. Our data is not yet powerful enough to break the degeneracy between different cosmological parameters, and therefore we only vary the amplitude of matter fluctuations, parametrised by $\sigma_8$. Our measurement thus corresponds to an independent constraint on the growth of structure at low redshifts, assuming that CMB data can reliably constrain all background evolution parameters ($\Omega_c$, $\Omega_b$, $H_0$, etc.). Combinations with other datasets (e.g. BAO measurements) may allow us to break these degeneracies independently from the CMB, but we leave this analysis for future work.

The resulting $68\%$ CL constraints on $\sigma_8$ are
\begin{equation}
  \sigmafid.
\end{equation}
The full marginalised distribution is shown in Figure \ref{fig:sigma_8}, together with the constraints from \planck \citep{Planck:2020:overview}, as well as the Kilo-Degree Survey \citep[KiDS, ][]{Heymans:2021}, and the Dark Energy Survey \citep[DES, ][]{Abbott:2022}. The fiducial measurement (first from the top in Fig. \ref{fig:sigma_8}) is in agreement with the weak lensing surveys, and at $1.2 \sigma$ difference from the CMB constraints by \planck. The measurement using linear matter power spectrum (second from the top in Figure \ref{fig:sigma_8}) is in agreement with the fiducial setup, which uses the {\tt HALOFIT} modelling, but it is closer than the {\tt HALOFIT} to the measurements from \planck. To test the robustness of this result to the choice of galaxy sample, we repeat the analysis for the fiducial flux and S/N cuts of \citetalias{Alonso:2020} ($2.0\,{\rm mJy}$, 5.0 respectively, third and fourth from the top in Figure \ref{fig:sigma_8}). We obtain $\sigmatest$, in agreement with the result found for the fiducial sample, but we note higher uncertainty of the estimations using this additional sample. These results are summarised in Table \ref{table:sigma_8}. The full posterior distribution of all model parameters is shown in Figure \ref{fig:corner}.

\subsection{Integrated Sachs-Wolfe effect}\label{ssec:results.isw}

\begin{figure}
    \centering
    \resizebox{\hsize}{!}{\includegraphics{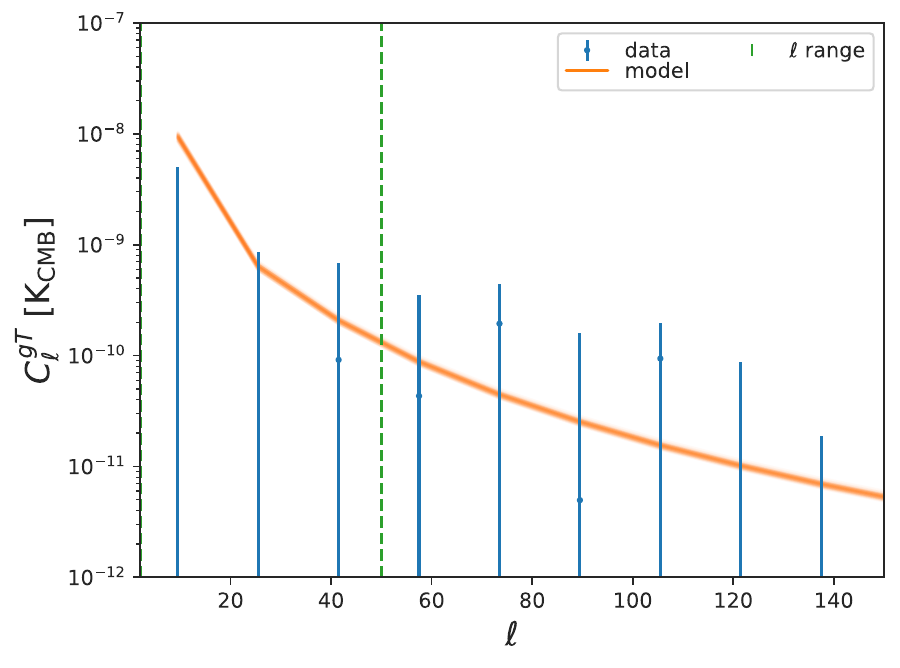}}
    \caption{Cross-correlation with the CMB temperature, based on fiducial bias constraints from the section \ref{ssec:results.bias}, using scales $2 < \ell < 50$. The signal-to-noise ratio is consistent with zero.}
    \label{fig:isw_fit}
\end{figure}

The ISW signal is especially useful for cosmology, as it is sensitive to the dark energy equation of state. However, sufficiently large sky coverage of the galaxy sample is needed for it to be detected, which is not yet the case for LoTSS. Figure \ref{fig:isw_fit} shows the measured cross-correlation between our LoTSS DR2 sample limited at $1.5\, {\rm mJy}$ and the CMB temperature anisotropies measured by \planck. This measurement was carried out using thinner $\ell$ bins ($\Delta\ell=16$) to concentrate on the largest scales, where the ISW signal is the most significant. The orange lines in the same plot show the theoretical prediction for values of the galaxy bias selected by the MCMC chains run in Section \ref{ssec:results.bias} for the $1/D(z)$ model using {\tt HALOFIT}. Fixing the galaxy bias to the best-fit value found in Section \ref{ssec:results.bias}, and comparing the $\chi^2$ value of the measured $C_\ell^{gT}$ with respect to the null-hypothesis and the best-fit model, we determine that the ISW signal is not significantly detected. A higher-significance measurement of this signal can be expected with future releases of LoTSS covering the full northern sky.

\section{Discussion}\label{sec:discussion}

We have shown that our results are in reasonable agreement with previous measurements of the galaxy bias for various radio samples, including previous CMB lensing cross-correlation analyses. The comparison with the real-space analysis of LoTSS DR2 carried out in \citetalias{Hale:2023} shows $1.3\sigma$ difference if we use only the auto-correlation and linear matter power spectrum, $1.8\sigma$ difference if we add the cross-correlation, and $2.3\sigma$ difference if we assume the {\tt HALOFIT} modelling for both approaches, which provides better fits in our case. As shown in \citet{Hamana:2022}, the difference in $S_8 = \Omega_m^{1/2}\sigma_8$ estimates between the real and harmonic space can be even larger than $1\sigma$. In our case, there are several possible reasons for the resulting difference.
\begin{itemize}
    \item The pure sample variance is due to the fact that both analyses actually do not use the same modes. We only expect harmonic space and real space methods to agree for full sky coverage or isotropic sampling of a statistically isotropic universe. The second aspect, the isotropic sampling, is indeed violated, and we tried to correct it by means of the weight mask and data cuts that we applied. However, the difference between the real and harmonic space analyses can point out that we did not correct for all the large-scale systematics.
    
    \item Both approaches treat the multi-component sources in different ways. In our case, it is a marginalisation over the amplitude of shot noise, whereas \citetalias{Hale:2023} selected the scales that allow for the effects resulting from the multi-component sources to be  avoided.

    \item The angular two-point correlation function can be affected by contamination from a dipole. \citet{Chen:2016} show that an excess of two-point correlation at the degree scale in the NVSS data set can be removed by properly removing the NVSS dipole before analysing the two-point correlation. A study of that issue in the scope of the LoTSS survey will be published in \citet{Bohme:2023}.
\end{itemize}

\section{Conclusion}\label{sec:conclusions}

We combined the LoTSS DR2 wide field and the LoTSS DR1 deep fields, supplemented by multi-wavelength data, with gravitational lensing from the \planck CMB to place constraints on the bias and its evolution for radio galaxies, and on the amplitude of matter perturbations. Our main results can be summarised as follows:
\begin{itemize}
  \item We obtain one of the most significant detections of the cross-correlation between radio and CMB lensing data, resulting in the S/N at a level of $\sigmaCgkhigh\sigma$.
  \item We show that the inclusion of CMB lensing information leads to a clear preference for an evolving galaxy bias, growing towards higher redshifts, as expected from linear theory. We determined that a linear bias evolution of the form $b_g(z)=b_{g,D}/D(z)$, where $D(z)$ is the linear growth factor, is able to consistently provide a good description of different sectors of the data. For a sample that is flux-limited at $1.5\,{\rm mJy}$, we measure $\biasd$, which at the median survey redshift provides $\biasval$. These results are also in good agreement with more flexible bias parametrisations (e.g.\ a quadratic polynomial in redshifts), which lead to similar constraints.
  \item Freeing up the value of $\sigma_8$, we were able to constrain it to $\sigmafid$ using our fiducial sample. The result is in good agreement with weak lensing surveys -- KiDS \citep{Asgari:2021, Heymans:2021} and DES \citep{Abbott:2022}, as well as CMB data from \planck.
  \item We attempted a first measurement of the ISW signal with LOFAR data, but found that the signal is compatible with zero.
\end{itemize}

Throughout this analysis, we used conservative scale cuts, $\ell < 250$ for $C_\ell^{gg}$, and $\ell < 500$ for $C_\ell^{g\kappa}$, and showed that for more permissive cuts ($\ell<(500,800)$), including mildly non-linear scales, the simple linear bias models used here are not able to fit the data adequately. More work is needed in order to provide a robust model for the bias of radio galaxies that extends to non-linear scales. This could be done by making use of perturbative bias expansions \citep{Matsubara:2008wx,Desjacques:2016bnm}, or phenomenological halo-based models \citep{Peacock:2000qk,Berlind:2001xk}. However, additional information, in the form of cross-correlations with other tracers (e.g. optical redshift surveys and tomographic cosmic shear data) will be necessary in order to disentangle a non-linear bias from evolutionary effects. 

The strength of our approach comes from the ability of high-resolution radio surveys to detect galaxies at high redshifts, and from the combination of both auto- and cross-correlations with CMB lensing, which allowed us to break degeneracies between the amplitude of matter perturbations and a galaxy bias, and to potentially constrain the redshift evolution of the latter. However, by far, the largest source of systematic uncertainty in our results is the lack of redshift information for radio continuum samples. We modelled and calibrated the redshift distribution of our sample using three deep fields within the LoTSS DR2 footprint. The resulting redshift distribution is still subject to caveats, due to the use of photometric redshifts, the small area covered by the deep fields, and the uncertainty due to radio sources with no optical cross-matches (around $5\%$ of our sample). In our analysis, however, we have propagated the $p(z)$ calibration uncertainties, by making the redshift distribution measurements part of the data vector, modelled together with the galaxy and CMB lensing power spectra. Future LoTSS data releases will include one additional deep field and even deeper observations of those fields, which will likely help reduce this source of uncertainty. 

The current LoTSS catalogue did not allow us to make a significant detection of the ISW effect. However, future data releases, covering the majority of northern sky, should allow us to improve this result. If included in the cosmological analysis, this measurement could help improve cosmological constraints, particularly in the context of dark energy, which is an important source of the ISW signal.

In this work, we have demonstrated the significant improvement on cosmological and astrophysical constraints from radio continuum data enabled by the inclusion of CMB lensing cross-correlations. This will allow future LOFAR data releases to start providing meaningful constraints on cosmological parameters, on par with (and in combination with) other probes of the large-scale structure.

\begin{acknowledgements}

SJN is supported by the US National Science Foundation (NSF) through grant AST-2108402, and the Polish National Science Centre through grant UMO-2018/31/N/ST9/03975.
DA acknowledges support from the Beecroft Trust, and from the Science and Technology Facilities Council through an Ernest Rutherford Fellowship, grant reference ST/P004474.
MBi is supported by the Polish National Science Center through grants no. 2020/38/E/ST9/00395, 2018/30/E/ST9/00698, 2018/31/G/ST9/03388 and 2020/39/B/ST9/03494.
DJS acknowledges support from the Bundesministerium f\"ur Bildung und Forschung (BMBF) ErUM-Pro grant 05A20PB1 and Ministerium f\"ur Kultur und Wissenschaft des Landes Nordrhein-Westfahlen Profilbildung 2020 grant B3D.
CLH acknowledges support from the Leverhulme Trust through an Early Career Research Fellowship.
AP is supported by the Polish National Science Centre grant UMO-2018/30/M/ST9/00757, and by COST (European Cooperation in Science and Technology) through grant COST Action CA21136 – “Addressing observational tensions in cosmology with systematics and fundamental physics (CosmoVerse)”.
AP and MBi acknowledge support from the Polish Ministry for Science and Higher Education through grant DIR/WK/2018/12.
CSH's work is funded by the Volkswagen Foundation.
CSH acknowledges additional support by the Deutsche Forschungsgemeinschaft (DFG, German Research Foundation) under Germany’s Excellence Strategy – EXC 2121 `Quantum Universe' – 390833306 and EXC 2181/1 - 390900948 (the Heidelberg STRUCTURES Excellence Cluster).
PT acknowledges the support of the RFIS grant (No. 12150410322) by the National Natural Science Foundation of China (NSFC).
JZ is supported by the project “NRW-Cluster for data intensive radio astronomy: Big Bang to Big Data (B3D)“ funded through the programme “Profilbildung 2020”, an initiative of the Ministry of Culture and Science of the State of North Rhine-Westphalia.
MBr acknowledges support from the Deutsche Forschungsgemeinschaft under Germany's Excellence Strategy - EXC 2121 “Quantum Universe” - 390833306 and from the BMBF ErUM-Pro grant 05A2023.
MJJ acknowledges support of the STFC consolidated grant [ST/S000488/1] and [ST/W000903/1] and from a UKRI Frontiers Research Grant [EP/X026639/1]. MJJ also acknowledges support from the Oxford Hintze Centre for Astrophysical Surveys which is funded through generous support from the Hintze Family Charitable Foundation.     

LOFAR is the Low Frequency Array designed and constructed by ASTRON. It has observing, data processing, and data storage facilities in several countries, which are owned by various parties (each with their own funding sources), and which are collectively operated by the ILT foundation under a joint scientific policy. The ILT resources have benefited from the following recent major funding sources: CNRS-INSU, Observatoire de Paris and Université d’Orléans, France; BMBF, MIWF-NRW, MPG, Germany; Science Foundation Ireland (SFI), Department of Business, Enterprise and Innovation (DBEI), Ireland; NWO, The Netherlands; The Science and Technology Facilities Council, UK; Ministry of Science and Higher Education, Poland; The Istituto Nazionale di Astrofisica (INAF), Italy.

This research was carried out using Python 3 \citep{python3:2009} and a number of software packages from which we enumerate the most significant ones for our analysis:
healpy \citep{Zonca:2019},
HEALPix \citep{Gorski:2005},
Astropy \citep{astropy:2013, astropy:2018, astropy:2022},
pymaster \citep{Alonso:2019},
pyccl \citep{Chisari:2019},
emcee \citep{emcee:2013},
getdist \citep{Lewis:2019},
NumPy \citep{NumPy:2020},
SciPy \citep{SciPy:2020},
IPython \citep{ipython:2007},
Pandas \citep{pandas:2010},
Matplotlib \citep{Matplotlib:2007},
seaborn \citep{seaborn:2021},
tqdm \citep{tqdm:2019}.

\end{acknowledgements}

\bibliographystyle{aa}
\bibliography{mybib}

\begin{appendix}

\section{Posterior probability distributions}\label{appendix:corner}

\begin{figure*}[h!]
    \centering
    \includegraphics[width=0.9\textwidth]{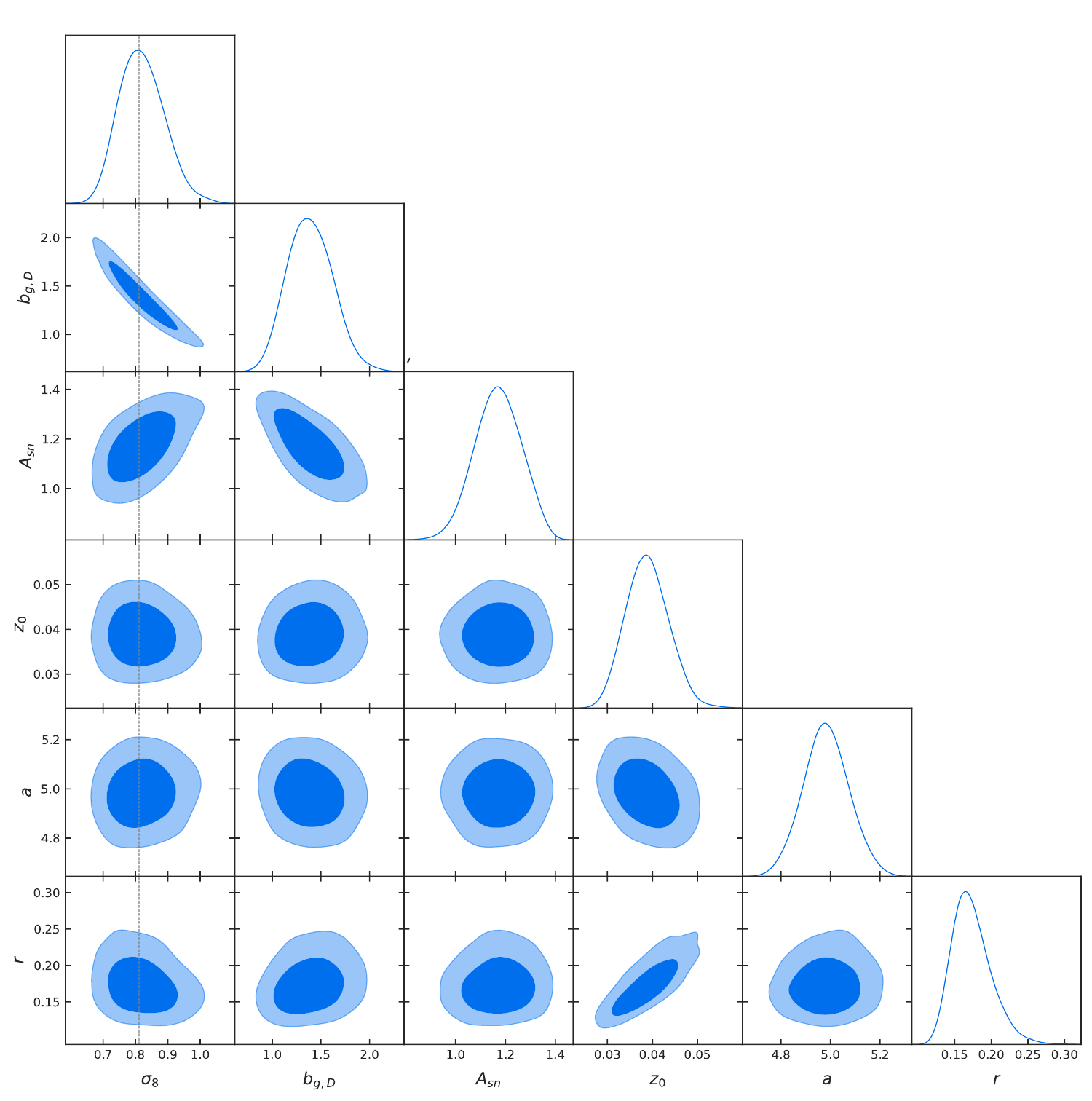}
    \caption{Probability density for constrained parameters, using flux density brighter than $2.0\,\rm{mJy}$, S/N higher than $5.0$, and the fiducial approach of $\ell < (250, 500)$, and {\tt HALOFIT} matter power spectrum. The vertical line marks the \planck constraints on $\sigma_8 = 0.81$.}
    \label{fig:corner_20}
\end{figure*}

Figs. \ref{fig:corner} and \ref{fig:corner_20} show the constraints on all model parameters for the 1.5 mJy and 2.0 mJy samples respectively.

\noindent\begin{minipage}{\textwidth}
    \centering
    \includegraphics[width=0.9\textwidth]{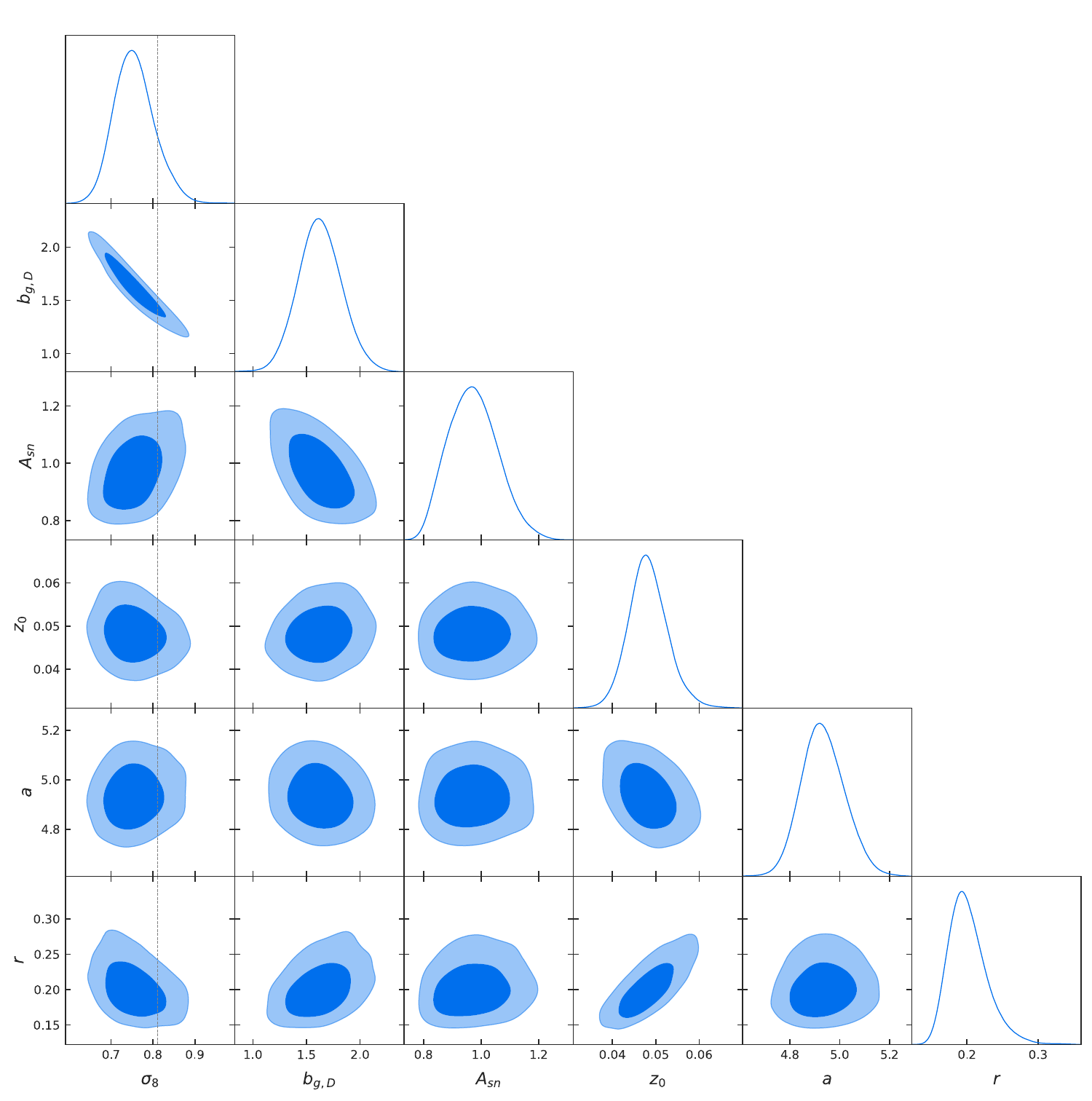}
    \captionof{figure}{Probability density for constrained parameters, using the fiducial approach of flux density brighter than $1.5\,\rm{mJy}$, S/N higher than $7.5$, $\ell < (250, 500)$, and {\tt HALOFIT} matter power spectrum. The vertical line marks the \planck constraints on $\sigma_8 = 0.81$.}
    \label{fig:corner}
\end{minipage}

\end{appendix}

\end{document}